\documentclass[fleqn,usenatbib]{mnras}

\usepackage[T1]{fontenc}
\usepackage{ae,aecompl}
\usepackage[normalem]{ulem}

\usepackage{graphicx}	% Including figure files
\usepackage{amsmath}	% Advanced maths commands
\usepackage{amssymb}	% Extra maths symbols
\usepackage{etoolbox}
\makeatletter
\patchcmd\@combinedblfloats{\box\@outputbox}{\unvbox\@outputbox}{}{%
   \errmessage{\noexpand\@combinedblfloats could not be patched}%
}%
 \makeatother
\newcommand{\code}[1]{\texttt{#1}}

\newcommand{\mesa}{\code{MESA}}

\newcommand{\nuclei}[2]{\ensuremath{\mathrm{^{#1}#2}}}

\newcommand{\carbon}[1][12]{\nuclei{#1}{C}}

\newcommand{\oxygen}[1][16]{\nuclei{#1}{O}}

\newcommand{\spr}{\mbox{$s$-process}}
\newcommand{\sprn}{\mbox{$s$ process}}

\newcommand{\rpr}{\mbox{$r$-process}}

\input{vectors}
\newcommand{\Sectff}[1]{{\ref{sec:#1}}}
\newcommand{\Sect}[1]{{\S\Sectff{#1}}}

\newcommand{\Eqref}[1]{{\ref{eq:#1}}}
\newcommand{\Eqff}[1]{{(\Eqref{#1})}}

\newcommand{\Eq}[1]{{Eq.~\Eqff{#1}}}

\newcommand{\fluid}[1]{$\mathcal{F}_#1$}

\newcommand{\ppmstar}{\textsc{PPMstar}}

\usepackage[usenames,dvipsnames]{color}
\usepackage[]{color}

\title[i-process yields of rapidly-accreting white dwarfs]{The i-process yields of rapidly-accreting white dwarfs from multicycle He-shell flash 
stellar evolution models with mixing parameterizations from 3D hydrodynamics simulations}

\author[P. A. Denissenkov et al.]{
Pavel A. Denissenkov,$^{1,2,\dagger}$\thanks{E-mail: pavelden@uvic.ca}
Falk Herwig,$^{1,2,\dagger}$
Paul Woodward,$^{2,3}$
Robert Andrassy,$^{1,2}$
\newauthor
Marco Pignatari$^{2,4,5,\dagger}$
and Samuel Jones$^{6,\dagger}$
\\
$^{1}$Department of Physics and Astronomy, University of Victoria, Victoria, BC, V8W 2Y2, Canada\\
$^{2}$Joint Institute for Nuclear Astrophysics - Center for the Evolution of the Elements, USA\\
$^{3}$LCSE and Department of Astronomy, University of Minnesota, Minneapolis, MN 55455, USA\\
$^{4}$E.A. Milne Centre for Astrophysics, Department of Physics \& Mathematics, University of Hull, HU6 7RX, United Kingdom\\
$^{5}$Konkoly Observatory, Research Centre for Astronomy and Earth Sciences, Hungarian Academy of Sciences, Konkoly-Thege,\\
 Mikl\'{o}s \'{u}t 15-17, H-1121 Budapest, Hungary\\
$^{6}$Computational Physics and Methods (CCS-2) and Center for Theoretical Astrophysics, Los Alamos National Laboratory,\\
 NM 87544, USA\\
$^\dagger$NuGrid Collaboration, \href{http://nugridstars.org}{http://nugridstars.org}\\}

\date{Accepted XXX. Received YYY; in original form ZZZ}

\pubyear{2018}

\begin{document}
\label{firstpage}
\pagerange{\pageref{firstpage}--\pageref{lastpage}}
\maketitle

% Abstract of the paper
\begin{abstract}
We have modelled the multicycle evolution of rapidly-accreting CO
white dwarfs (RAWDs) with stable H burning intermittent with strong
He-shell flashes on their surfaces for $0.7\leq M_\mathrm{RAWD}/M_\odot\leq 0.75$ 
and [Fe/H] ranging from $0$ to
$-2.6$.  We have also computed the i-process nucleosynthesis yields
for these models. The i process occurs when convection driven by the
He-shell flash ingests protons from the accreted H-rich surface layer,
which results in maximum neutron densities $N_\mathrm{n,max}\approx
10^{13}$\,--\,$10^{15}\ \mathrm{cm}^{-3}$.  The H-ingestion rate and
the convective boundary mixing (CBM) parameter $f_\mathrm{top}$ adopted in
the one-dimensional nucleosynthesis and stellar evolution models are
constrained through 3D hydrodynamic simulations. 
The mass ingestion rate and, for the first time, the scaling laws 
for the CBM parameter $f_\mathrm{top}$ have been determined from 3D hydrodynamic simulations.
We confirm our previous result that the high-metallicity RAWDs have a low mass
retention efficiency ($\eta \la 10\%$). A new
result is that RAWDs with [Fe/H]\,$\la -2$ have $\eta\ga 20\%$,
therefore their masses may reach the Chandrasekhar limit and they may eventually explode as SNeIa. This result and the
good fits of the i-process yields from the metal-poor RAWDs to the
observed chemical composition of the CEMP-r/s stars suggest that some
of the present-day CEMP-r/s stars could be former distant members of
triple systems, orbiting close binary systems with RAWDs that may have
later exploded as SNeIa.
\end{abstract}

% Select between one and six entries from the list of approved keywords.
% Don't make up new ones.

%%%%%%%%%%%%%%%%%%%%%%%%%%%%%%%%%%%%%%%%%%%%%%%%%%

%%%%%%%%%%%%%%%%% BODY OF PAPER %%%%%%%%%%%%%%%%%%

\section{Introduction}

The type Ia supernovae (SNeIa) are thermonuclear explosions of carbon-oxygen (CO) white dwarfs (WDs)
\citep[e.g.][]{hoyle60,Hillebrandt2000,Hillebrandt2013,Churazov2014,Livio2018}. In the single degenerate (SD) channel of SNIa progenitors,
that was originally proposed by \cite{Schatzman1963} and \cite{Whelan1973}, it is assumed
that the CO WD is a primary star of a close binary system with a main-sequence, subgiant or red-giant-branch
companion. The secondary star fills its Roche lobe and donates matter from an H-rich envelope to the WD, and, as a result,
the WD will explode when its growing mass $M_\mathrm{WD}$ approaches the Chandrasekhar limit $M_\mathrm{Ch}\approx 1.38 M_\odot$.
Initially, the primary star of such a binary system was an intermediate-mass star with $M\approx 2.5$\,--\,$7 M_\odot$, the upper
boundary of this mass interval depending on the amount of convective boundary mixing and 
C-burning rate \citep{Chen2014}. It left a core --- the CO WD --- after having lost the rest of its mass during
a common-envelope event, when it arrived at the asymptotic-giant branch (AGB) and filled its Roche lobe. 
Given that the CO cores of AGB stars can grow in mass only up to $\sim 1 M_\odot$ \citep{Chen2014}, the SD channel can work only if
the accreted H-rich matter is first processed into He and then into C and O, while being efficiently retained on the WD.

Figure 9 of \cite{Nomoto1982}  summarizes the results of the previous investigations of H accretion
onto CO WDs at different rates $\dot{M}_\mathrm{acc}$ (an update of this figure can be found in \citealt{Nomoto2007}). 
It shows that stable burning of accreted H occurs in a very narrow interval of 
$\dot{M}_\mathrm{acc}$ around a value of $\sim 10^{-7} M_\odot\ \mathrm{yr}^{-1}$ that linearly increases
with $M_\mathrm{WD}$. At the lower rates, H is processed into He via thermal flashes that become stronger
when $\dot{M}_\mathrm{acc}$ decreases, eventually leading to thermonuclear runaways typical for the classical novae.
At the higher rates, the non-processed H accumulates in an expanding envelope, so that such a rapidly accreting
WD would be resembling a red giant.

However, an important aspect of the evolution of rapidly accreting white dwarfs is the fact 
that even if accreted H is burning stably the He shell will ignite in a thermonuclear runaway 
when enough He has been accumulated as H-shell burning ash, just as 
in a thermal-pulse AGB star \citep{Cassisi1998}. 
A few consecutive He-shell flashes at the end of H accretion were computed by \cite{Idan2013}.
Following \cite{Nomoto1982}, various outcomes of pure He accretion onto CO WDs at different values of $\dot{M}_\mathrm{acc}$
have been studied in a number of papers \citep[e.g.][]{Piersanti2014,Wang2015}.
The data obtained in such simulations are used in models of binary population synthesis to estimate
a theoretical SNIa rate for the SD channel, simply assuming that the He accretion rate matches that of H
in the regime of stable H burning \citep[e.g.][]{Han2004}. 

Recently, \citet[][hereafter Paper I]{Denissenkov2017a} have presented
the results of the first stellar evolution computations in which rapid
accretion and stable burning of H on CO WDs were repeatedly
interrupted by strong He-shell flashes that were followed by mass loss
caused either by the super-Eddington luminosity wind or by the
common-envelope event resulting from the expansion of the WD envelope
overflowing its Roche lobe, after which the H accretion
resumed. These simulations show a low efficiency of He
retention ($\eta_\mathrm{He}\la 10\%$) and, because all the processes
that accompany the rapid H accretion were taken into account, they
provide estimates of the low retention efficiency of the total
accreted mass ($\eta\approx \eta_\mathrm{He}$) by the
rapidly-accreting WDs (RAWDs)\footnote{We define an RAWD as a WD that
  accretes H rapidly enough for its stable burning to be maintained on
  the WD surface. This definition is slightly different from the one
  used by \cite{Lepo2013} whose RAWDs can have the higher accretion
  rates and lose mass via the optically thick wind.}.  In one of the
models even a negative value of $\eta_\mathrm{He}$ was found meaning
that $M_\mathrm{WD}$ was decreasing with time. Given that the binary
population synthesis models predict an order of magnitude lower SNIa
rates for the SD channel when optimistically assuming that
$\eta_\mathrm{He} = 100\%$, this result of Paper I makes the SD
channel highly unlikely. Fortunately, there are a number of
alternative channels of SNIa progenitors, various pros and cons of
which are discussed by \cite{Wang2018} and \cite{Livio2018}.

Paper I has proposed a new application for the former SD channel,
namely, instead of growing in mass towards $M_\mathrm{Ch}$ and
exploding as SNeIa, the RAWDs could be a stellar site of the
intermediate (i) process of neutron captures by heavy isotopes
\citep{Cowan1977}. This idea is based on the similarity of physics
that one encounters in the RAWD models and in the model of the
post-AGB star Sakurai's object (V4334 Sagittarii) in which the
elemental abundance signature of the i-process nucleosynthesis was
observed by \cite{Asplund1999} and interpreted by
\cite{Herwig2011}. Indeed, in both cases the He-shell burning drives
convection that ingests protons from an H-rich envelope. This triggers
the reactions $^{12}$C(p,$\gamma)^{13}$N, $^{13}$N(e$^+\nu)^{13}$C and
$^{13}$C($\alpha$,n)$^{16}$O, the first one taking place close to the
top and the third one near the bottom of the He convective zone, and
$^{13}$N decaying into $^{13}$C while being transported downwards.

The exact mechanism of how this convective flow operates in
different cases is still under investigation. Sakurai's object is
believed to be a very late thermal pulse, i.e. the He-shell flash
occured when the H-burning shell of the star had already turned off
and the star had evolved around the knee\footnote{A segment of a WD cooling track where
the evolution with nearly constant luminosity and increasing effective temperature
changes to the evolution with both of them decreasing.} in the HRD. Stellar
evolution models predict an immediate split of the He-shell flash
convection zone which would prevent any mixing of protons and the
burning products of p capture on \carbon\ into the hottest bottom
region of the pulse-driven convection zone. \citet{Herwig2014} have
shown that in three-dimensional hydrodynamic simulations the H
ingestion triggers a Global Oscillation of Shell H-ingestion (GOSH)
which drastically rearranges the structure of the He-shell flash
convection zone. So far it has not been possible to follow the
long-term evolution of this event past the first GOSH. The
one-dimensional, spherically symmetric nucleosynthesis simulations
of \citet{Herwig2011} adopted the approach that mixing between the
upper layer in which protons and \carbon\ react and the bottom layer
in the pulse-driven convection zone, where neutrons can be released
via the $\carbon[13](\alpha,n)\oxygen[16]$ reaction, continues until
the observed chemical composition of Sakurai's object has been reproduced as well as possible. 
This \emph{delayed-split} scenario is not yet fully supported by the initial
three-dimensional hydrodynamics simulation results by
\citet{Herwig2014}. The initial GOSH in those simulations happens at
a time when the amount of protons consumed is still insufficient to
explain the neutron exposure required to explain observations, and 
the evolution past the initial GOSH in three dimensions remains unclear.

With an ongoing supply of protons, most of which are transformed into
neutrons in the above reactions, the neutron number density at the
bottom of the He zone can reach a value of $N_\mathrm{n}\sim
10^{15}\ \mathrm{cm}^{-3}$ intermediate between the values
characteristic of the $s$ ($N_\mathrm{n}\la 10^{11}\ \mathrm{cm}^{-3}$)
and $r$ ($N_\mathrm{n}\ga 10^{20}\ \mathrm{cm}^{-3}$)
process.
Very-late thermal pulse objects such as the post-AGB star
Sakurai's object will only experience one H-ingestion event. Their
nucleosynthesis production is very unique, but the impact on a
galactic chemical evolution scale is neglible. RAWDs on the other
hand continuously accrete from a close binary companion, and can
potentially experience dozens of He-shell flashes followed by
mass-loss episodes before the concomitant changes in the binary system
parameters terminate the rapid H accretion. Along with their low mass
retention efficiency, this makes the RAWDs a potentially important
galactic sources of heavy elements, with distinct elemental and
isotopic abundance signatures different from those produced in the $s$
and $r$ process.  For example, i process from RAWDs can make a significant
contribution to the first n-capture peak of the solar system abundance
distribution, as demonstrated by combining RAWD i-process yields with stellar
population synthesis and galactic chemical evolution models
\citep{Cote2018}.

RAWDs are not the only possible sites of i-process nucleosynthesis
triggered by H ingestion into a He convective zone. The other sites
can be low-metallicity and low-mass thermally-pulsing AGB stars \citep{Iwamoto2004,Lugaro2012},
He-core flash in low-matallicity RGB stars \citep{Campbell2010},
super-AGB stars \citep{Jones2016}, and H- and He-burning shell merger
in Population-III massive stars \citep{Clarkson2018,Banerjee2018}. 

In Paper I, we have considered the rapid
($\dot{M}_\mathrm{acc}=1$--$2\times 10^{-7}
M_\odot\,\mathrm{yr}^{-1}$) accretion of only solar-composition matter
onto CO WDs with the masses $0.65 M_\odot$, $0.73 M_\odot$ and $1
M_\odot$. The present work extends the set of our RAWD models to
sub-solar metallicities, while keeping their masses close to
$M_\mathrm{WD}\approx 0.73 M_\odot$. The main goals of this paper are
to describe the methods that we use to simulate the multicycle
evolution of RAWDs (Section~\ref{sec:evol_meth}) and the i-process
nucleosynthesis in their He-flash convection zones during H ingestion
(Section~\ref{sec:ipr_meth}), and to present the results of our new
computations of the RAWD evolution and i-process yields for a range of metallicity
(Section~\ref{sec:results}) that have been used in
\cite{Cote2018}. Section~\ref{sec:hydro} describes the 3D hydrodynamic
simulations that support our 1D estimates of H mass ingestion rates
for the RAWD models. Section~\ref{sec:concl} concludes the paper.

% Example figure
\begin{figure*}
  \includegraphics[width=144mm]{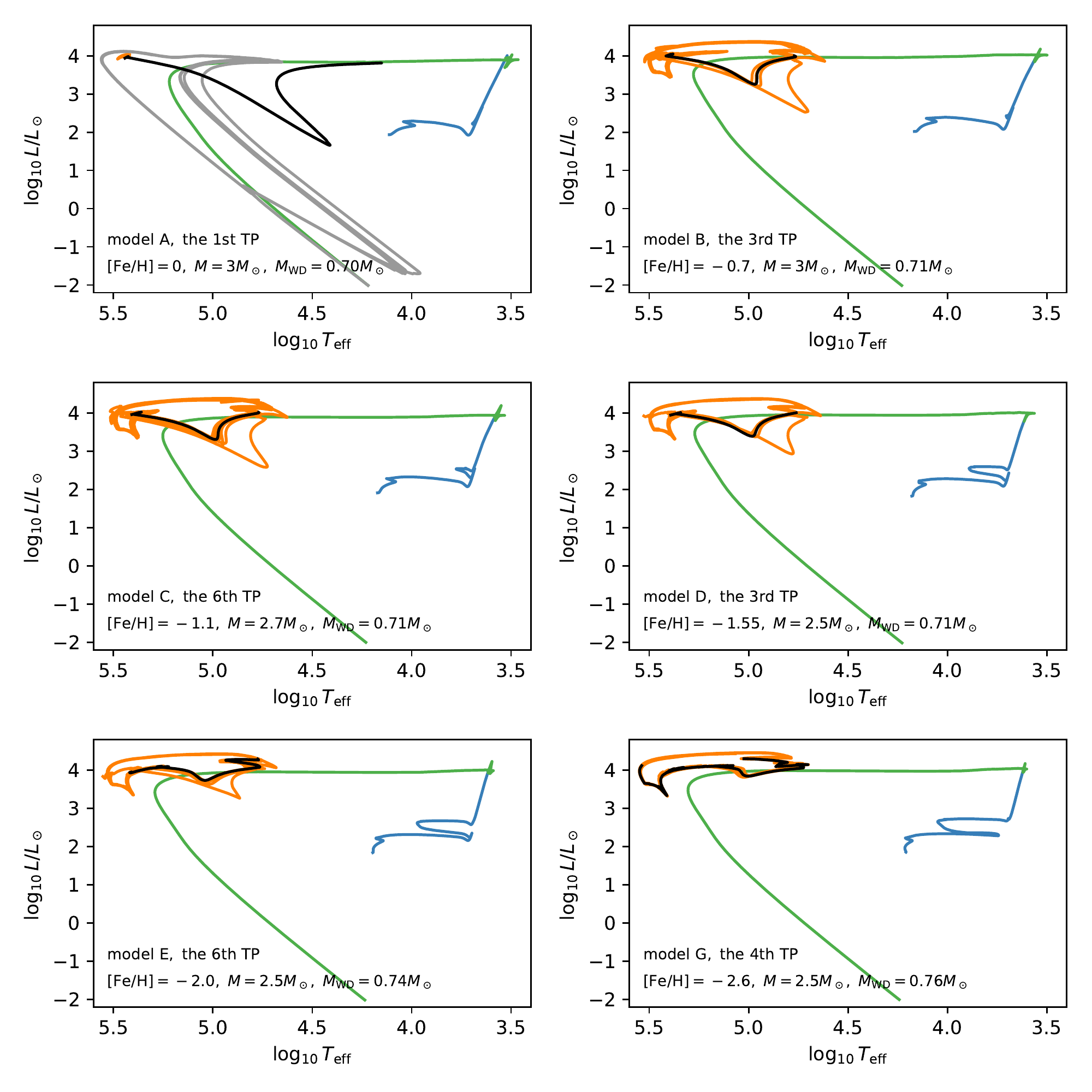}
  \caption{The evolutionary tracks of the progenitors of our RAWD
    models (blue and green, the pre-MS phase is not shown) and the tracks of the multicycle RAWD
    evolution (orange; the evolution goes counterclockwise in the cycles). The grey curve in model A panel shows the
    nova-like evolutionary loops that are used to reach the starting
    point, at the high-$T_\mathrm{eff}$ and high-$L$ knee of the nova
    loop, of the RAWD evolution modelling. Text in the panels
    indicates the models and the serial numbers of their He-shell
    thermal pulses (TP) that are chosen for the post-processing
    simulations of the i-process nucleosynthesis, the black curves
    showing their tracks.  The other parameters of these models are
    listed in Table~\ref{tab:models}. Model F is not shown here because
    its track and i-process yields are very similar to those of model G
    that we use to reproduce the surface abundances of heavy elements in CEMP-r/s stars.}
    \label{fig:fig1}
\end{figure*}

\section{Methods used for one-dimensional stellar evolution and nucleosynthesis simulations}

\subsection{The RAWD evolution}
\label{sec:evol_meth}

The new RAWD models are computed with the revision 7624 of the \mesa\
stellar evolution code \citep{Paxton2011,Paxton2013}.  We use the
reaction rates from the {\em JINA Reaclib} database \citep{Cyburt2010}
and the \mesa\ default equation of state.  The nuclear network includes
31 species, from neutron to $^{28}$Si, that can participate in 60
reactions of the pp chains, four CNO cycles, NeNa and MgAl cycles, as
well as the He (triple-$\alpha$, $^{12}$C($\alpha,\gamma)^{16}$O, $^{16}$O($\alpha,\gamma)^{20}$Ne, $^{20}$Ne($\alpha,\gamma)^{24}$Mg,
$^{13}$C($\alpha$,n)$^{16}$O, and other $(\alpha,\gamma)$, $(\alpha$,n) and $(\alpha$,p)
reactions with the included isotopes) and C burning. The initial mixtures of elements and
isotopes are prepared using the solar-system chemical composition of
\cite{Asplund2009} that is scaled to specified values of
     [Fe/H]\footnote{[A/B] $=\log_{10}(N(\mathrm{A})/N(\mathrm{B}))-
       \log_{10}(N_\odot(\mathrm{A})/N_\odot(\mathrm{B}))$, where
       $N(\mathrm{A})$ and $N(\mathrm{B})$ are the abundances (number
       densities or mass fractions) of the nuclides A and B.}. We
     assume that for [Fe/H]\,$\leq -0.7$ the initial mixtures are
     $\alpha$-element enhanced with [$\alpha$/Fe]\,$=+0.4$. The
     appropriate Type 1 and Type 2 (with enhanced C and O abundances)
     OPAL and low-temperature molecular opacities have been prepared
     for such mixtures \citep{Denissenkov2017b} and used in our
     computations.

The CO WD models are made with the inlists from the \mesa\
test suite example {\tt make\_co\_wd}.
With these inlists the \mesa\ code first computes the
evolution of an intermediate-mass star from the pre-main sequence to
the completion of the first He-shell thermal pulse on the AGB (blue
curves in Figure~\ref{fig:fig1}, initial masses given in each
panel), then the Bl\"{o}cker AGB wind parameter
\citep{Bloecker1995} is increased from $0.1$ to $5$ to mimic the
enhanced mass loss caused by a common-envelope interaction in a close
binary system in which the AGB star overflows its Roche
lobe. As a result, the model star leaves the AGB and evolves towards
and down the WD cooling track (green curves in the same Figure).

Our computations include convective boundary mixing (CBM) adopting the
exponentially decaying diffusion model \citep{Herwig1997}:
\begin{eqnarray}
D(r) = D_0\exp\left(\frac{-2|r-r_0|}{f H_P}\right),
\label{eq:DOS}
\end{eqnarray}
where $D_0$ is a value of the convective diffusion coefficient
provided by the mixing-length theory (MLT) and $H_P$ is the pressure scale height,
both evaluated at $r=r_0$ in the vicinity of the respective convective
boundary.  For the boundaries of the H and He convective cores we have
adopted the value of $f = 0.014$ that is close to the one constrained
by the position of the terminal-age main sequence (TAMS) in a large number of
stellar clusters \citep{Herwig2000}.  For the top and bottom
boundaries of the He-flash convective zone we use the values of
$f_\mathrm{top} = 0.1$ and $f_\mathrm{bot} = 0.008$ that are equal or
close to those obtained in the multi-dimensional hydrodynamic
simulations of the He-shell flash convection by
\cite{Herwig2007}. 
The value for $f_\mathrm{bot}$ is consistent with a
number of abundance observables of \spr\ elements and
H-deficient stars \citep{Herwig:2005jn,Werner:2006bf,Battino:2016bna}.
Our stellar models evolve through the phases of
convective H and He core burning and on the RAWD phase they experience
convective He shell burning, therefore the above constraints for
the values of $f$ can be applied to them.
We provide further support from new
hydrodynamic simulations for our choice of $f_\mathrm{top}$ in \Sect{hydro}. 

When the CO WD model cools down to $\log_{10} L/L_\odot = -2$, we initiate a slow accretion of H-rich matter on it with 
$\dot{M}_\mathrm{acc}\approx 10^{-8} M_\odot\,\mathrm{yr}^{-1}$ using the \mesa\ mass-change control parameter
in a new inlist with the same as before input physics. A higher value of $\dot{M}_\mathrm{acc}$ at this stage
would result in \mesa\ iterations having not converged. We assume that the accreted matter has the initial
chemical composition of the binary. Because the accretion rate is now lower than the one required for stable H burning,
our model exhibits mild H-shell flashes, each of them being followed by the expansion of its envelope. To stop
the model from becoming a red giant, we enforce a mass loss by implementing the \mesa\ super-Eddington wind
presription with an artificially reduced value of $L_\mathrm{Edd}$, 
so that the model returns to the accretion phase and continues to make nova-like loops
on the Hertzsprung-Russel diagram (a grey curve in the top-left panel of Figure~\ref{fig:fig1};
it is shown only for the progenitor of the solar-metallicity RAWD model A). During the stable
H burning at a higher accretion rate a RAWD remains near the high-$T_\mathrm{eff}$ and high-$L$ ``knee'' of
a nova loop \citep[e.g.][]{Wolf2013}, therefore we switch the accretion rate to a higher value 
in a model that is located near the knee. We adjust a value for $\dot{M}_\mathrm{acc}$ that would guarantee 
stable H burning and allow relatively large time steps between consecutive evolutionary models.

\begin{figure*}
  \includegraphics[width=144mm]{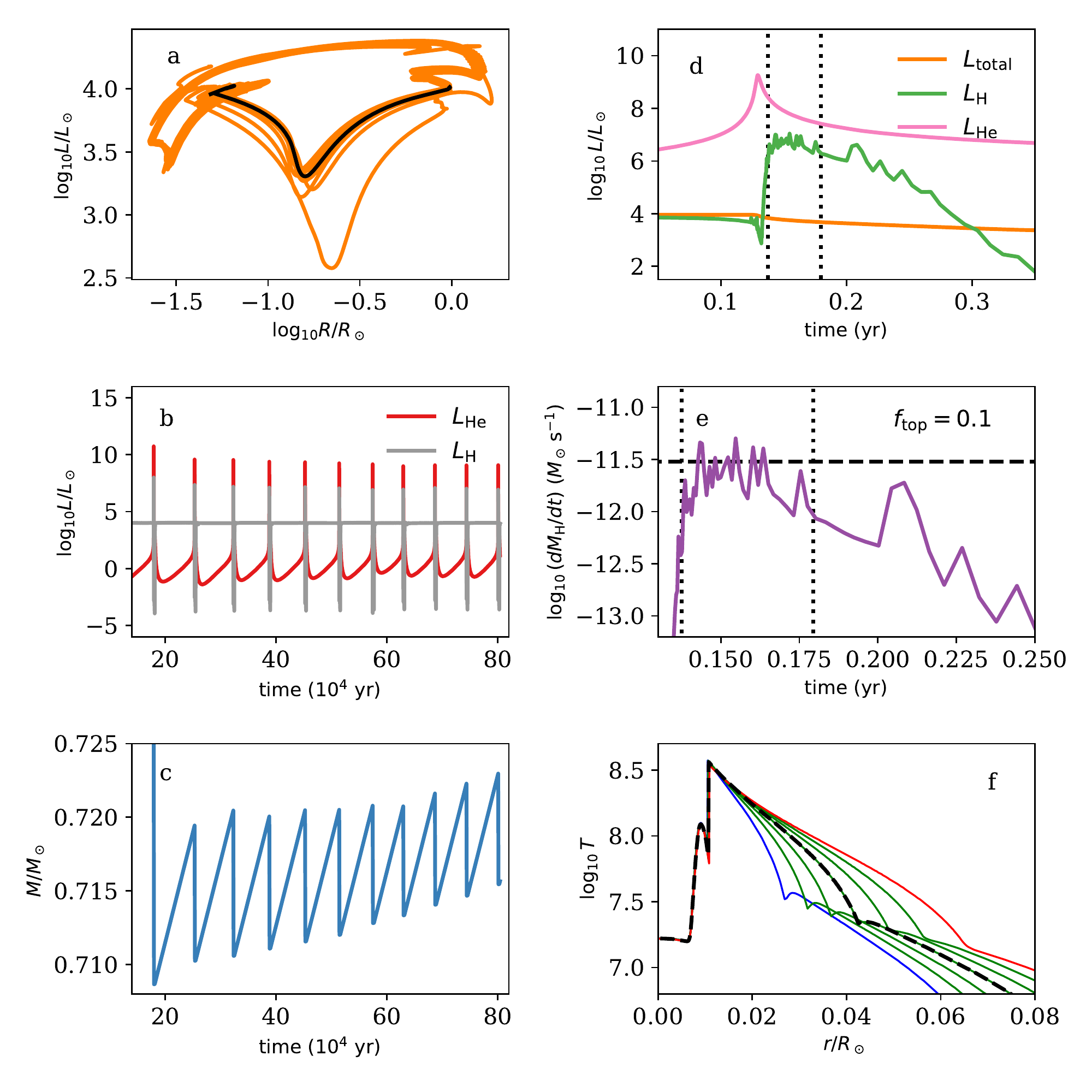}
  \caption{The evolution of luminosities, total mass, H-ingestion rate and temperature
    profile for the RAWD model C from the middle-left panel of Figure~\ref{fig:fig1}. 
    Panel a shows the changes of the total luminosity with the radius during eleven consecutive
    He shell flashes (the evolution goes counterclockwise in the cycles) for which the corresponding changes of the H and He luminosities
    are shown in panel b and the changes of the total mass in panel c. The expansion of
    the RAWD is bounded by its assumed $2\,M_\odot$ Roche-lobe radius (Equation \ref{eq:myRL}).
    Panels d and e show how the luminosities and H-ingestion rate change with time
    during the 6th He shell flash (the black line segment in panel a). The wiggles on the $L_\mathrm{H}$
    and $\dot{M}_\mathrm{H}$ curves are caused by the fact that the CBM prescription (\ref{eq:DOS}) is
    non-local and therefore its implementation in \mesa\ is decoupled from the solution of stellar structure 
    equations. Panel f: the evolution of the temperature profile in the RAWD during the 6th He shell flash. The black dashed
    curve was used in the post-processing computations of nucleosynthesis in the He
    convective zone.
    }
    \label{fig:fig2}
\end{figure*}

The RAWDs spend most of their time stably burning the accreted H at
the knees of nova loops, during which they should be seen as
super-soft X-ray sources, unless being obscured by the ejected
circum-binary matter \citep{vandenHeuvel1992,Lepo2013,Woods2016}.
Just as in thermal-pulse AGB stars, when a critical mass of He
is accreted from the H-burning shell, a He-shell flash occurs causing
an expansion of the accreted envelope (Figures~\ref{fig:fig2}a and
\ref{fig:fig2}f).  Whereas the mass loss via the radiation-driven
super-Eddington wind becomes less efficient at sub-solar
metallicities, a fast and efficient mass loss can still be assumed for
a star in a close binary system when it expands and overflows its
Roche lobe.  Therefore, a slightly modified \mesa\ scheme for the
Roche-lobe wind has been implemented in the present work to model the
mass loss by our RAWD models during their expansion driven by the
He-shell flashes.  While the \mesa\ code approximates the Roche-lobe
mass-loss rate as
$$
\dot{M}_\mathrm{RL} = \dot{M}_\mathrm{RL,0} \exp\left(\frac{R-R_\mathrm{RL}}{H_\mathrm{RL}}\right),
$$
where $\dot{M}_\mathrm{RL,0}$ is a base value for which we use $10^{-3} M_\odot\,\mathrm{yr}^{-1}$, 
$R_\mathrm{RL}$ is the Roche-lobe radius of the mass-losing star,
and $H_\mathrm{RL}$ is the Roche-lobe wind scale height, we prefer to use the following prescription:
\begin{eqnarray}
\dot{M}_\mathrm{RL} = \dot{M}_\mathrm{RL,0} \left(\frac{R}{R_\mathrm{RL}}\right)^6
\label{eq:myRL}
\end{eqnarray}
for $R > 0.2R_\mathrm{RL}$. The implementation of our prescription for
the Roche-lobe mass loss, that is equivalent to the \mesa\ approximation
for $|R - R_\mathrm{RL}|\ll R_\mathrm{RL}$ if one uses $H_\mathrm{RL}
= R_\mathrm{RL}/6$, results in a smoother evolution of the RAWD
models.  After having lost a certain amount of mass with the
Roche-lobe wind, which determines the mass retention efficiency, our
RAWD model returns to the knee, and the H mass accretion resumes.

\subsection{The RAWD i-process nucleosynthesis}
\label{sec:ipr_meth}

The i process in an RAWD commences when the top of the He convective
zone reaches the bottom of its H-rich surface layer, soon after the
He-shell flash peak luminosity.  At this moment, the He-shell
convection begins to ingest protons (the left dotted line in Figure~\ref{fig:fig2}d
marks the moment when the H luminosity suddenly increases by more than
two orders of magnitude as a result of the beginning H ingestion).
It should be noted that our
\mesa\ computations find such H ingestion in all of our RAWD
models even when $f_\mathrm{top} = 0$
(Paper~I), i.e. even when convective boundary mixing at the top boundary of
the He convective zone is not included (although the H ingestion rate
$\dot{M}_\mathrm{H}$ does positively correlate with
$f_\mathrm{top}$).

The i process in our RAWD models is simulated in post-processing
computations similar to those carried out to model the
i-process nucleosynthesis in Sakurai's object
\citep{Herwig2011,Denissenkov2018}.  These computations use the NuGrid
multi-zone post-processing nucleosynthesis parallel code mppnp
\citep{Pignatari2016} customized for the H-ingestion He-shell problem.
The input data for this code include a static or time-dependent
structure of the He convective zone (this work uses the first option), i.e. the
radius $r$, temperature $T$, density $\rho$, and convective diffusion
coefficient $D_\mathrm{conv}$ at each point of its mass mesh, the
chemical compositions of the He zone and of the
ingested matter, the mass ingestion rate $\dot{M}_\mathrm{ing}$ and
its duration $t_\mathrm{ing}$.

Whereas the ingested matter 
has the initial chemical composition of the binary, the
composition of the He convective zone at the beginning of H ingestion
is obtained by processing the initial
mixture through complete H burning followed by its processing via
partial He burning, until the increasing C abundance matches its value
from the corresponding \mesa\ RAWD model, for which we use
the NuGrid single-zone code ppn
\citep[cf.][]{Denissenkov2018}.

Our i-process nucleosynthesis simulations include $\sim 1000$ isotopes and $\sim 15000$ reactions.
The reaction rates for these simulations are taken from the same list
of references as in \cite{Denissenkov2018}.
We adopt an equally spaced 100-zone mass grid for the He shell
region by interpolating the stellar structure variables to the new mesh. At each time step
$\Delta t$, we add $X_\mathrm{o}^k\dot{M}_\mathrm{ing}\Delta t$ mass
of the $k$th isotope from the envelope to the top
$\Delta M = 1$\,--\,$4\times 10^{-4} M_\odot$ of the He
shell that occupies $\sim 10$ mass zones, as described in
Appendix~\ref{app:mdot}.

\subsection{The mass ingestion rate and duration}

The mass ingestion rate $\dot{M}_\mathrm{ing}$ is
determined from a combination of constraints from three-dimensional
simulations (see \Sect{hydro}) and the \mesa\ RAWD stellar evolution simulations.
Given the modelling choices described in the previous section the 1D
stellar evolution simulations predict the ingestion of H-rich envelope
material into the He-shell flash convection zone as it expands outward
in Lagrangian coordinate. The protons ingested into the convection
zone lead to a H-burning luminosity $L_\mathrm{H}$ which reflects the
mass ingestion rate through the relation $\dot{M}_\mathrm{H} =
X_\mathrm{surf}\dot{M}_\mathrm{ing}$, where $X_\mathrm{surf}$ is the H
mass fraction at the RAWD surface, and
\begin{eqnarray}
\dot{M}_\mathrm{H}\approx \frac{L_\mathrm{H}}{\varepsilon_\mathrm{H}},
\label{eq:dMHdt}
\end{eqnarray}
where $\varepsilon_\mathrm{H}$ is the energy released per one gram of
burned H.

Because only the first two reactions of the CNO cycle are fast enough
to occur in the He-shell convective zone during a convective overturn
time, we assume that $\varepsilon_\mathrm{H} =
0.667\,\varepsilon_\mathrm{CNO}$, where
$\varepsilon_\mathrm{CNO}\approx 6.3\times 10^{18}\, \mathrm{erg\,g}^{-1}$ 
is the energy released per one gram of H transformed into
He in the full CNO cycle, and the factor $0.667$ is the fraction of
this energy produced per one gram of consumed H in the reactions
$^{12}$C(p,$\gamma)^{13}$N and $^{13}$N(e$^+\nu)^{13}$C.

For the duration of mass ingestion $t_\mathrm{ing}$, we adopt the values
estimated from our \mesa\ RAWD models. We use $\dot{M}_\mathrm{H}(t)$ curves, 
like the one shown in Figure~\ref{fig:fig2}e, to estimate a time interval $t_\mathrm{ing}$
during which $\dot{M}_\mathrm{H}$ remains close to its maximum value wiggling around some nearly constant mean value,
the latter giving us an estimate of $\dot{M}_\mathrm{H}$. 
We check that the estimated values of $t_\mathrm{ing}$ do not
exceed their upper limits constrained by the condition $t_\mathrm{ing}
< M_\mathrm{env}/\dot{M}_\mathrm{ing}$. 
The adopted values of $\dot{M}_\mathrm{ing}$ and $t_\mathrm{ing}$ used
in the different simulations are summarized in Table~\ref{tab:models}.
Although this is a somewhat subjective method of
choosing the $\dot{M}_\mathrm{ing}$ and $t_\mathrm{ing}$ input data for the post-processing nucleosynthesis simulations,
we think its accuracy is consistent with our using a static temperature profile,
like the black dashed curve in Figure~\ref{fig:fig2}f, in these simulations.

\begin{table*}
	\centering
	\caption{Summary of the one-dimensional RAWD simulation
          parameters ($L_\mathrm{He}^\mathrm{ing}$ is the He luminosity at the beginning of H ingestion). 
            }
	\label{tab:models}
	\begin{tabular}{clccccccc} % four columns, alignment for each
		\hline
		model & [Fe/H] & $M_\mathrm{WD}\,(M_\odot)$ & $\dot{M}_\mathrm{acc}\,(M_\odot\,\mathrm{yr}^{-1})$ & 
                $\log_{10}(L_\mathrm{He}^\mathrm{max}/L_\odot)$ & 
                $\log_{10}(L_\mathrm{He}^\mathrm{ing}/L_\odot)$  & 
                $\dot{M}_\mathrm{ing}\,(M_\odot\,\mathrm{s}^{-1})$ & 
                $t_\mathrm{ing}\ \mathrm{(yr)}$ & $\eta\,(\%)$ \\
		\hline
		A & $0.0$ & $0.70$ & $2.6\times 10^{-7}$ & $10.9$ & $9.1$  & $2.2(35)\times 10^{-12}$ & $0.17(0.024)$  & -- \\
		B & $-0.7$ & $0.71$ & $1.7\times 10^{-7}$ & $9.5$ & $8.5$  & $2.0\times 10^{-12}$ & $0.054$ & $4.9$\\
		C & $-1.1$ & $0.71$ & $1.5\times 10^{-7}$ & $9.3$ & $8.4$  & $4.0\times 10^{-12}$ & $0.042$ & $4.9$ \\
		D & $-1.55$ & $0.71$ & $ 1.5\times 10^{-7}$ & $9.3$ & $8.5$  & $4.2\times 10^{-12}$ & $0.083$ & $9.6$ \\
		E & $-2.0$ & $0.74$ & $1.7\times 10^{-7}$ & $8.7$ & $8.1$  & $3.3\times 10^{-12}$ & $0.060$ & $27$ \\
		F & $-2.3$ & $0.75$ & $1.5\times 10^{-7}$ & $9.2$ & $8.6$  & $2.4\times 10^{-11}$ & $0.058$ & $19$ \\
		G & $-2.6$ & $0.75$ & $1.5\times 10^{-7}$ & $8.5$ & $8.0$  & $6.7\times 10^{-12}$ & $0.087$ & $29$ \\
		\hline
	\end{tabular}
\end{table*}

\section{Ingestion rates and convective boundary mixing parameters from 3D hydrodynamic simulations}
\label{sec:hydro}

The ingestion of material from the stable layer into the
convection zone is the result of complex mixing processes at the
convective boundary that may involve global, large-scale flow modes
revealed in full $4\pi$ three-dimensional hydrodynamic simulations
\citep{Woodward2015}. As mentioned in \Sect{evol_meth} we adopt in
one-dimensional simulations the exponentially decaying CBM model
with an efficiency parameter $f_\mathrm{top}$ to describe this mixing.

Both the mass ingestion rate and the convective boundary parameter
$f_\mathrm{top}$ can be determined from hydrodynamic simulations, as
demonstrated by \citet{Jones2017}. For this purpose we have performed
3D hydrodynamic simulations of the He-shell flash convection zone and
H ingestion in a RAWD using the \ppmstar{} code \citep{Woodward2015}.
It is an explicit Cartesian-grid-based code for 3D hydrodynamics built around the
Piecewise-Parabolic Method \citep[PPM;][]{woodward_colella81,
woodward_colella84, colella_woodward84, woodward86, woodward06}. The code
advects the fractional volume of the lighter fluid in a two-fluid scheme using
the Piecewise-Parabolic Boltzmann method \citep[PPB;][]{woodward86, Woodward2015}.
Thanks to PPB's use of subcell information, it needs two to three times fewer
grid cells along all three axes than PPM to reach the same level of fidelity
in the advection of a quantity, like the multifluid mixing fraction,
whose value is conserved along stream lines. The code was designed with strong
emphasis on parallel efficiency and it has performed past simulations of shell
convection on up to 440{,}000 CPU cores on the NCSA Blue Waters computer
\citep{Woodward2015,Herwig2014}.

The radial stratification of simulations is based on model A from
Paper~I. 
Like model A in the present paper, model A 
in Paper~I had the solar initial chemical composition, but it had a bit lower mass of $0.65\,M_\odot$. 
We consider the point in time $7.44\mathrm{hr}$  after the
beginning of the second He-shell flash in that model, when its He luminosity has
dropped to $4.10 \times 10^9\,L_\odot$ from its maximum value of $7.4
\times 10^{10}\,L_\odot$, and H ingestion into the He shell has just
started.

The initial stratification of the 3D simulations follows the same
approach as in \cite{Woodward2015} and approximates the 1D model with
three polytropes: a lower stable layer (radial range $6\,\mathrm{Mm} <
r < 7.4\,\mathrm{Mm}$), a convection zone ($7.4\,\mathrm{Mm} < r <
33.5\,\mathrm{Mm}$), and an upper stable layer ($33.5\,\mathrm{Mm} < r
< 50.0\,\mathrm{Mm}$). We neglect radiation pressure, which
contributes less than $25\%$ to the total pressure in the 1D model,
and we use the equation of state for a monoatomic ideal gas. To obtain
a similar overall stratification with the slightly different equation
of state, we use a small mean molecular weight $\mu_1 = 0.3$ for the
fluid \fluid{1} initially filling the upper stable layer.  The rest of
the simulation domain contains fluid
\fluid{2} with $\mu_2 = 1.4$. The two fluids are allowed to react with
each other via the $^{12}$C(p,\,$\gamma$)$^{13}$N reaction, assuming
that \fluid{1} contains 88.6\% of protons and \fluid{2} contains
20.4\% of $^{12}$C by number.  The subsequent decay of $^{13}$N is not
considered. Convection in the He shell is driven by volume heating
applied between the radii $7.9$\,Mm and $8.9$\,Mm.

The 3D simulations are done in $4\pi$ geometry on a Cartesian grid. We
measure the luminosity dependence of the mass ingestion rate using
runs E8, E13, and E15 (see Table~\ref{tab:hydro_sims}), which cover a
range of $1.4$\,dex in the driving luminosity $L_\mathrm{He}$ at the
grid resolution of $768^3$, and one high-resolution run E10 ($1536^3$)
with the same driving luminosity as its $768^3$ equivalent E8.

Hydrogen ingestion starts as soon as the first upwelling plumes reach
the upper convective boundary. After a few convective overturns a
balance between hydrogen ingestion and burning is reached and the
convective-reactive flow becomes quasi-stationary. We do not find a
GOSH in these H-ingesion simulations which differs from the results
found in the case of Sakurai's object \citep{Herwig2014}. This
corresponds to the result from 1D stellar evolution reported in
Paper~I and also found in most simulations here that the H ingesition
does not cause a split of the convection zone, with the exception of
model A where a split happens in very late phases (see
\Sect{results}).

\begin{figure}
  \includegraphics[width=\columnwidth]{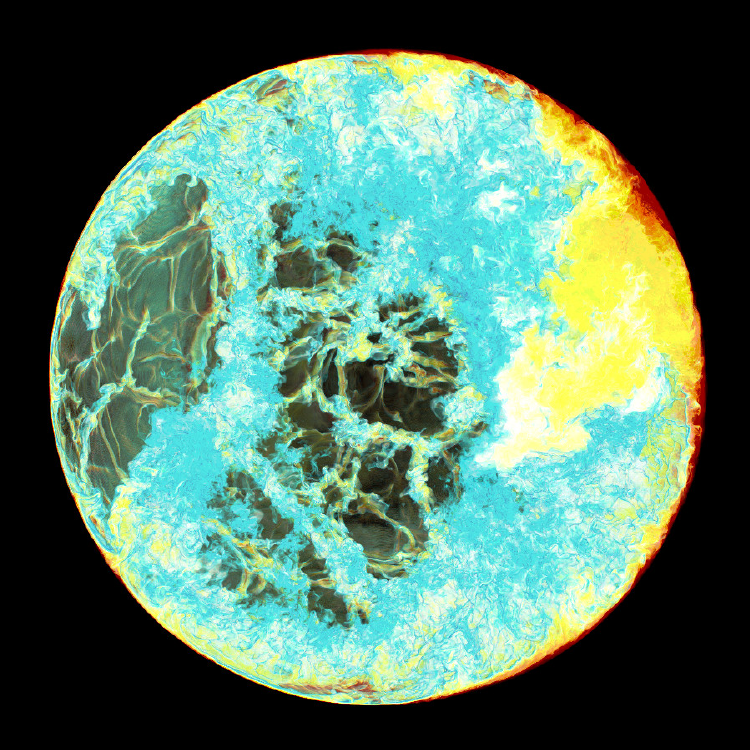}
  \caption{Fractional volume (FV) of the fluid ingested into the convection zone in
  run E10 at $t = 359.8$\,min. 
  The colour scheme is such that the lowest values of FV are transparent, 
  then when FV increases the colour changes from blue through white and yellow to red.
  Finally, the black colour corresponds to FV = 1, so that the stable layer is not seen.
  %The colour scale is logarithmic and very low concentrations are transparent. 
  The front half of the sphere is not shown and
  the camera is looking into the back half of the sphere in this rendering.}
  \label{fig:fig3}
\end{figure}

Figure~\ref{fig:fig3} shows that the ingestion
process is dominated by large scales. The average hydrogen luminosity
$L_\mathrm{H}$ reaches $2$--$3$\% of the driving luminosity $L_\mathrm{He}$
(Table~\ref{tab:hydro_sims}). Ingestion events localised in time and
space can have a much stronger influence on the flow than these small
values suggest, but they are not strong enough to launch a global
ingestion instability such as the GOSH phenomenon observed by
\citet{Herwig2014} in their 3D simulations of Sakurai's object.

\begin{figure}
  \includegraphics[width=\columnwidth]{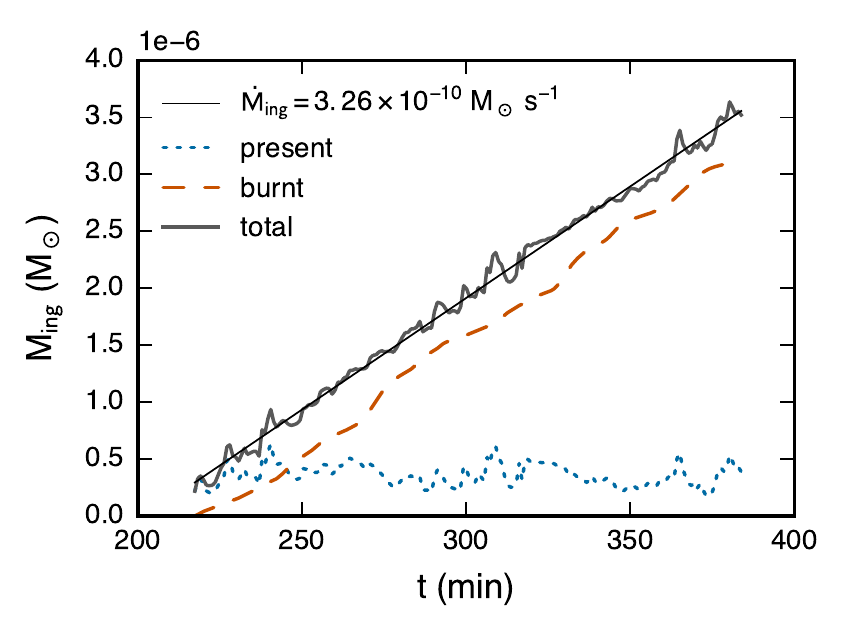}
  \caption{Time dependence of the amounts of ingested fluid present and burnt in
  the convection zone in run E10. The rate of change of their total mass is the
  ingestion rate $\dot{M}_\mathrm{ing}$. The amount of fluid burnt is counted
  from zero at the beginning of the shown time series.} \label{fig:fig4}
\end{figure}

The amount of mass $M_\mathrm{ing}(t)$ ingested into the convection zone by a
time $t$ is the sum of mass $M_\mathrm{p}(t)$ present in the convection zone at
this time and mass $M_\mathrm{b}(t)$ burnt in the convection zone by this time.
The radius $r_\mathrm{ub}$ of the upper boundary of the convection zone
increases in time as a result of both mass ingestion and thermal expansion. We
define $r_\mathrm{ub}$ to be the radius at which the radial gradient of the root-mean-square
horizontal velocity $v_\mathrm{h}(r)$ reaches a local maximum and we integrate
the density of the ingested fluid up to the radius $r_\mathrm{top} =
r_\mathrm{ub} - H_{v,\mathrm{ub}}$, where the velocity scale height
$H_{v,\mathrm{ub}} = (\partial v_\mathrm{h} / \partial r)^{-1}$ is evaluated at
$r_\mathrm{ub}$. The subtraction of $H_{v,\mathrm{ub}}$ mitigates issues
related to the large contrast in the concentration of the ingested fluid
between the boundary region and the bulk of the convection zone
\citep[for details, see][]{Jones2017}. To find the burnt mass $M_\mathrm{b}(t)$, we
compute the mass burning rate from spherically-averaged profiles of
temperature, density, and fractional volume of ingested fluid at
regularly-spaced points in time; $M_\mathrm{b}(t)$ is then obtained by time
integration. The resulting time dependence of $M_\mathrm{p}$, $M_\mathrm{b}$,
and $M_\mathrm{ing}$ in run E10 is shown in Fig.~\ref{fig:fig4}. We obtain
the ingestion rate $\dot{M}_\mathrm{ing}$ by fitting a straight line to
$M_\mathrm{ing}(t)$.
\begin{table*}
  \centering
  \caption{Summary of our 3D \ppmstar{} simulations.}
  \label{tab:hydro_sims}
  \begin{tabular}{ccccc}
    \hline
    run & grid & $\log_{10}L_\mathrm{He}/L_\odot$ &
    $L_\mathrm{H}/L_\mathrm{He}$ &
    $\dot{M}_\mathrm{ing}\,(M_\odot\,\mathrm{s}^{-1})$ \\
    \hline
    E8   & $768^3$  & $9.95$  & $0.025$ & $2.32 \times 10^{-10}$ \\
    E10  & $1536^3$ & $9.95$  & $0.034$ & $3.26 \times 10^{-10}$ \\
    E13  & $768^3$  & $10.65$ & $0.019$ & $9.45 \times 10^{-10}$ \\
    E15  & $768^3$  & $11.35$ & $0.020$ & $5.04 \times 10^{-9}$  \\
    \hline
  \end{tabular}
\end{table*}

\begin{figure}
  \includegraphics[width=\columnwidth]{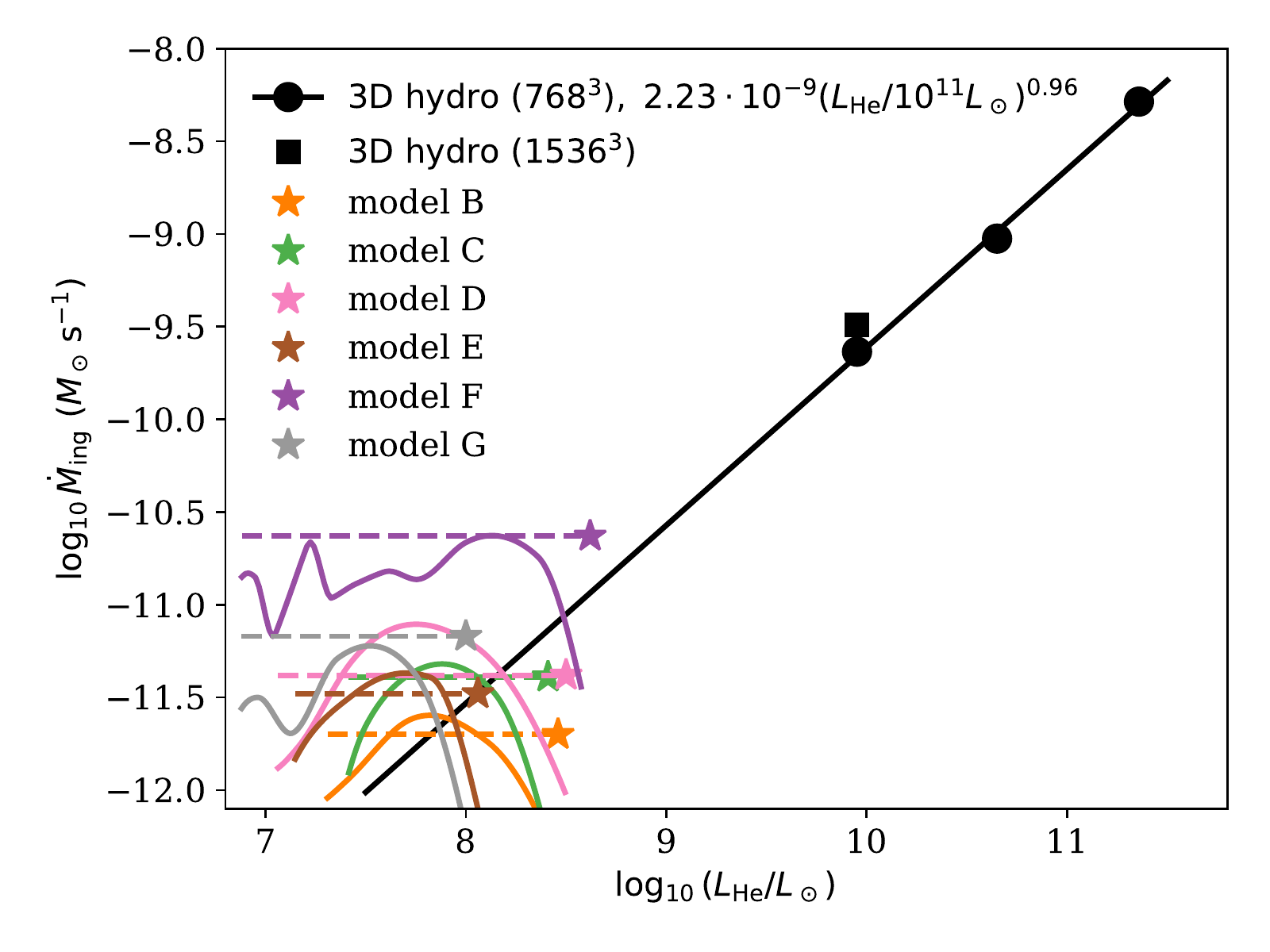}
  \caption{Ingestion rates as functions of He-burning lumiosity
      that is driving convection according to the RAWD stellar
      evolution models and 3D hydrodynamic simulations. The black
      line is the $L_\mathrm{He}$-scaling relation for the mass
      ingestion rate from the 3D hydrodynamic simulations that used the
      RAWD model A from Paper~I for the initial setup.  The filled black
      circles are the results of the individual $(768)^3$ hydro
      simulations and the filled black square is the result of the $(1536)^3$ hydro run (Table~\ref{tab:hydro_sims}).
      The star symbols with adjacent horizontal dashed line segments are the average mass ingestion rates estimated for our 1D RAWD
      models as described in Section~\ref{sec:yields}, while the arc curves of the same colours
      show smooth fits to the actual variations of $\dot{M}_\mathrm{ing}$ with $L_\mathrm{He}$ in these models.
      The star symbols are located at $L_\mathrm{He} = L_\mathrm{He}^\mathrm{ing}$ that corresponds to
      the beginning of  H ingestions in 1D models (Table~\ref{tab:models})}
    \label{fig:fig5}
\end{figure}
Figure~\ref{fig:fig5} shows that the ingestion rate scales in
proportion to the driving luminosity $L_\mathrm{He}$, in agreement
with the results of \citet{Jones2017} on the O-shell convection in
massive stars. This is likely to be caused by the fact that the
ingestion rate is limited by the amount of work needed to be done to
overcome the buoyancy of the ingested material, as argued by
\citet{Spruit2015}. The ingestion rate in run E8 ($768^3$) is only
$26\%$ lower than in run E10 ($1536^3$) of the same luminosity, which
provides an idea of the resolution dependence of these entrainment
rate results.

The ingestion rates measured in 1D models B--F are close to the
scaling relation established by the 3D hydrodynamic simulations, which
may not have been expected given the numerous differences between the
1D and 3D models. For the RAWD models in Figure~\ref{fig:fig5}, we
have used the He-shell luminosity at the moment when the top of the He
convective zone reaches the bottom of the H-rich envelope and, as a
result of this, the H-burning luminosity quickly increases.
These luminosities are somewhat
lower than their corresponding peak He luminosities (e.g., panel d in
Figure~\ref{fig:fig2}). In any case, the ingestion rates adopted for
our nucleosynthesis simulations are consistent with the results
obtained with the 3D hydrodynamic simulation in the sense that they
are close to the scaling relation between the driving luminosity of
the convection and the ingestion rate established by three $768$-grid
hydrodynamic simulations.

\begin{figure}
  \includegraphics[width=\columnwidth]{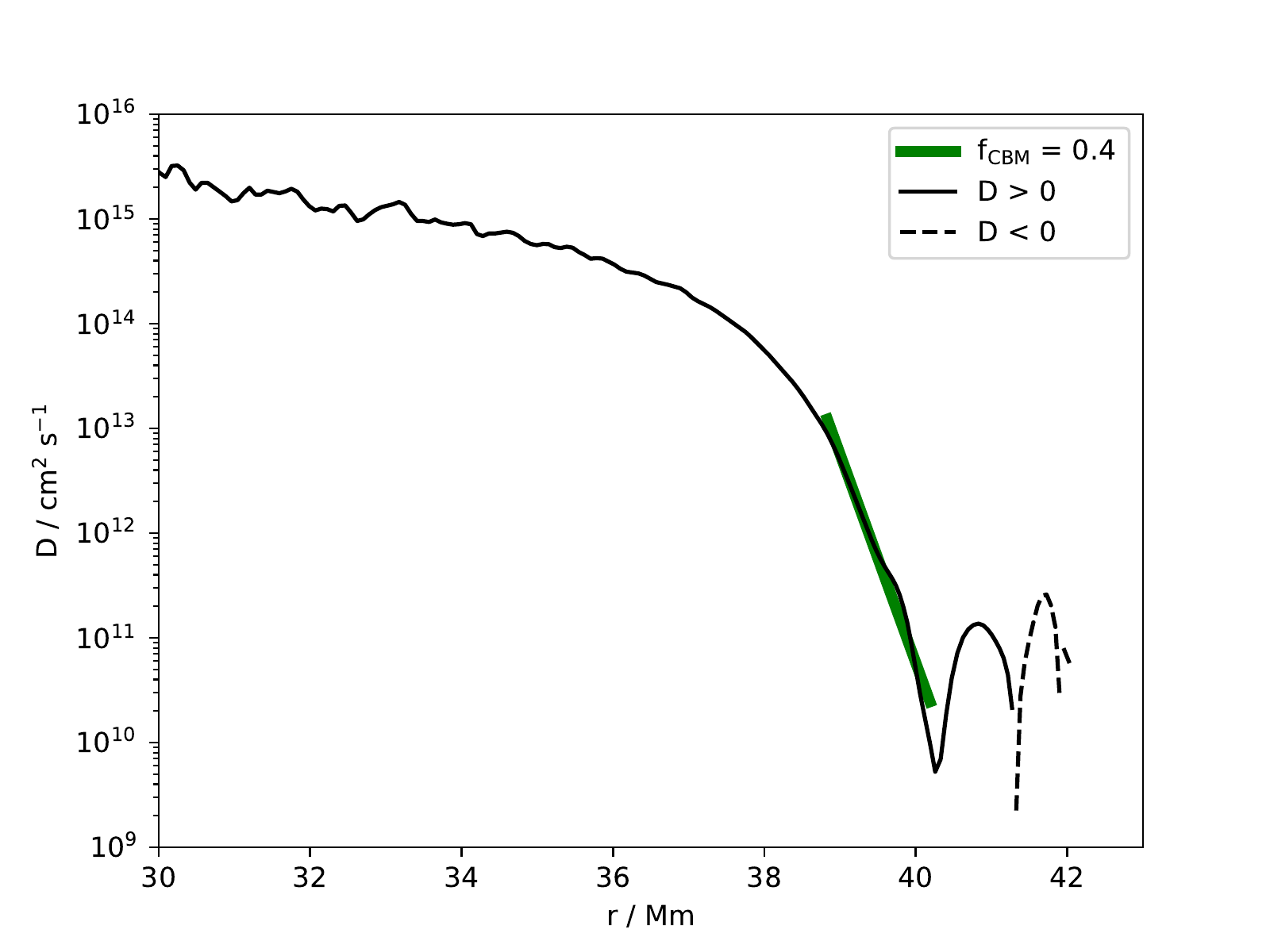}
  \caption{
    Determination of the $f_\mathrm{top}$ mixing parameter from the
    high-grid-resolution 3D hydrodynamics simulation E10.}
    \label{fig:fig6}
\end{figure}
\begin{figure}
  \includegraphics[width=\columnwidth]{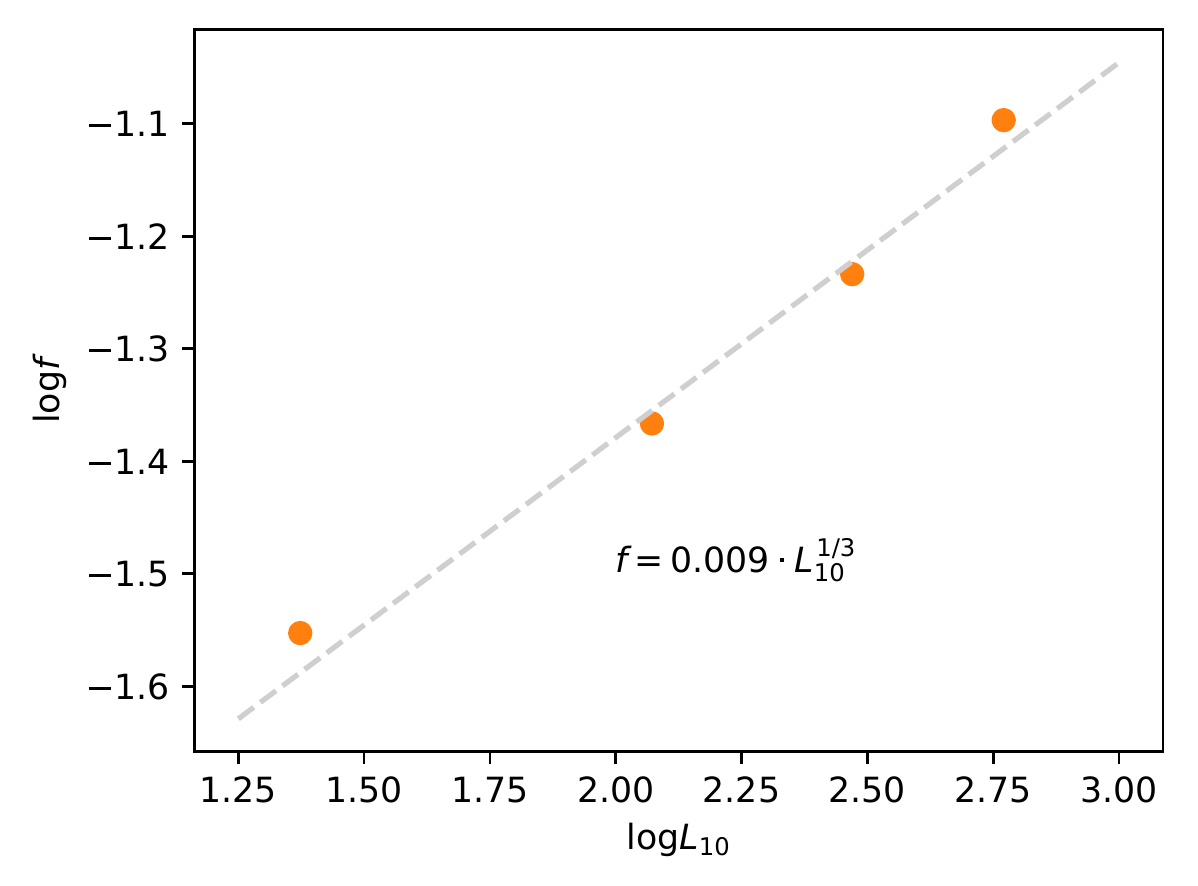}
  \caption{Scaling relation of the convective boundary mixing
    parameter $f_\mathrm{top}$ versus the driving luminosity based on
    a series of 1536-grid simulations of O-shell convection with the
    same setup as in \citet{Jones2017}.}
    \label{fig:fig7}
\end{figure}
We can also obtain information on the convective boundary mixing
efficiency at the top of the He-shell flash convection zone. As in
\citet{Jones2017} we determine the $f$ parameter from the
evolution of the spherically averaged radial profiles of the abundance
of the H-rich fluid that is entrained into the He-shell flash
convection zone (Figure~\ref{fig:fig6}). This is done by solving the
inverted Lagrangian diffusion equation which gives the diffusion coefficient
profile that would have been needed to advance from an
abundance profile at a time $t$ to a profile at time
$t + \Delta t$ by means of a 1D diffusion
process. The time difference $\Delta t$ is usually taken to be one or a
few convective overturning time scales. The details of this procedure have been improved somewhat
over the approach in \citet{Jones2017} and will be described elsewhere
in detail. For the high-resolution E10 simulation we determine
$f_\mathrm{top} = 0.434$.

This value is larger than the one
appropriate for our 1D simulations because the E10 hydrodynamic
simulation has been performed at a driving luminosity that is 141 times
higher than, for example, the He-burning lumninosity in the stellar
evolution run C ($\log L_\mathrm{He} = 7.8$). In order to scale the
$f$ value obtained in our higher-luminosity hydro simulation to the
actual lower luminosity of the stellar evolution RAWD model we use the
scaling relationship
\begin{equation}
  f_\mathrm{top} \propto L_\mathrm{drive}^{1/3}
  \label{eq:fscaling}
\end{equation}
between the driving luminisity of a shell convection and the
convective boundary mixing parameter $f_\mathrm{top}$ at the top of
the convection zone. This relationship has been derived from a series
of new $1536$-grid 3D hydrodynamic simulations of O-shell convection
as in \citet{Jones2017}, but with an improved version of the
\ppmstar{} code (Figure~\ref{fig:fig7}). These simulations, the $f$
determination and the resulting scaling law will be described in
detail elsewhere. However, this relationship can be motivated within a
simplistic picture in which the CBM $f$ parameter is a measure of how
much convective plumes can deform the convective boundary and
penetrated into it before being decelerated by their negative buoyancy
upon entering the stably stratified regions. It would be the momentum
of the convective plume that determines the level of penetration and
boundary deformation. The convective velocity
scales with one third power with the luminosity
\citep{biermann32,porter:00a,muller_janka15,Jones2017}. The scaling
relation \Eq{fscaling} reflects this reasoning: $f \propto v \propto
L^{1/3}$.

Applying  relation \Eq{fscaling} to scale
$f_\mathrm{top} = 0.434$ from the
E10 driving luminosity of $\log L_\mathrm{He} = 9.95$ to the driving
luminosity of the stellar evolution model C at the time of H ingestion
($\log L_\mathrm{He} = 7.8$) we obtain $f_\mathrm{top} = 0.08$. This
provides strong support for the value $f_\mathrm{top} = 0.10$ that we
have adopted in the RAWD stellar evolution simulation which are all at
a similar He-burning luminosity at the time of ingestion.

\section{Results}
\label{sec:results}

\subsection{The RAWD multicycle evolution}

Using the methods described in Section~\ref{sec:evol_meth}, we have
computed the evolution of seven RAWD models with the metallicities and
WD masses listed in Table~\ref{tab:models} along with other model
parameters.  The evolutionary tracks of six of these models are shown in
Figure~\ref{fig:fig1}, where their initial masses are also
indicated. Except the solar-metallicity model A, we have simulated the
RAWD evolution for many cycles, typically more than five,
(orange curves in
Figure~\ref{fig:fig1}). The multicycle evolution of the
solar-metallicity RAWD models was discussed in Paper~I.  Black curve
segments in Figure~\ref{fig:fig1} highlight the relatively
short-lasting evolutionary phase of the He-shell thermal pulse (TP)
whose serial number is specified for each model and that has been
chosen for the i-process post-processing nucleosynthesis computations
in Section~\ref{sec:yields}.

% Example table
All of the RAWD models have nearly the same initial central temperature with $\log_{10}T_\mathrm{c}\approx 7.2$ and use the same
WD Roche-lobe radius $R_\mathrm{RL,WD} = 2 R_\odot$ in Equation~\ref{eq:myRL} that corresponds to the orbital period $P\approx 1.2$ days for
a secondary mass of $\sim 2 M_\odot$.

\begin{figure}
  \includegraphics[width=\columnwidth]{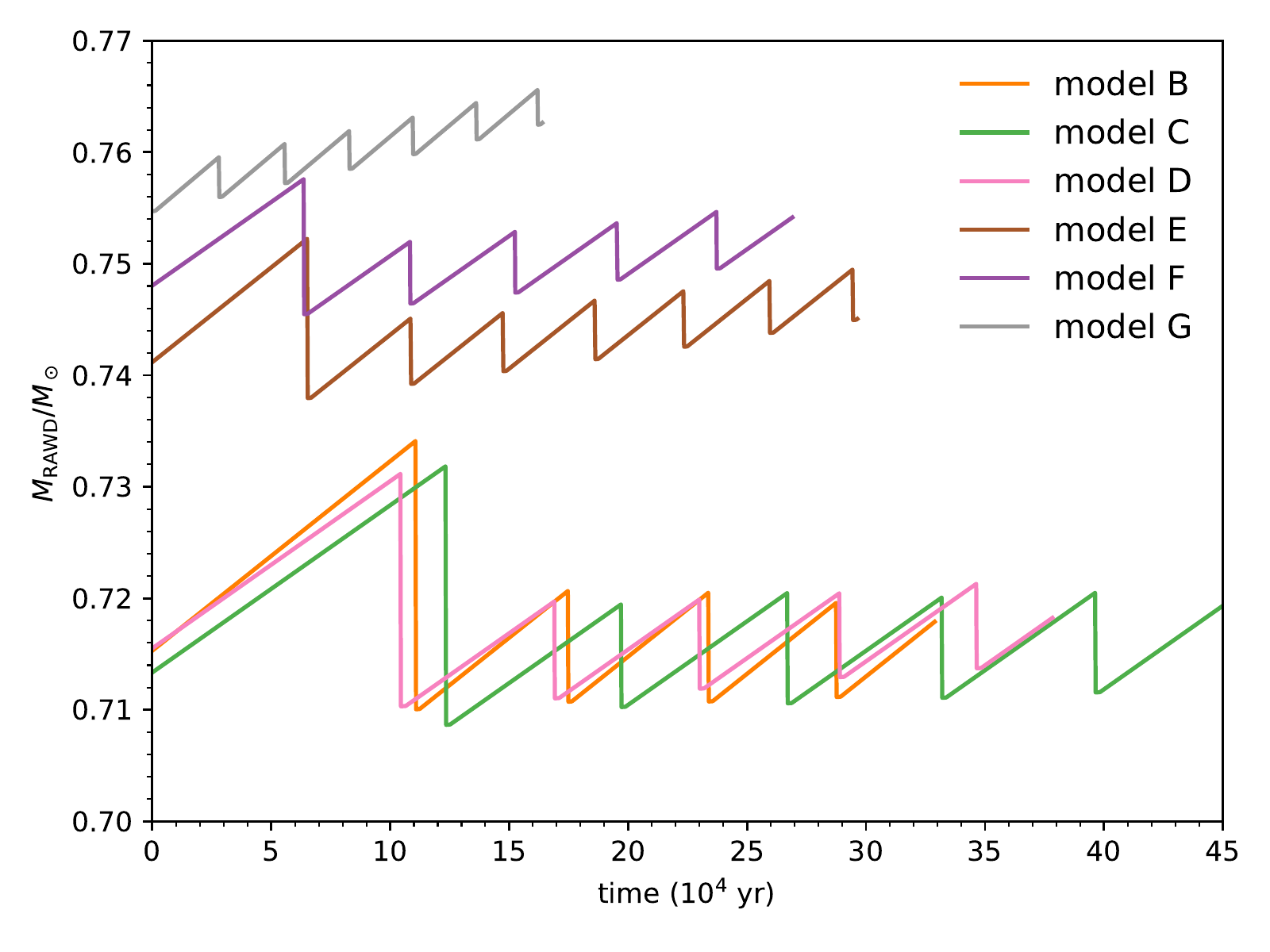}
  \caption{The changes of the total mass of our RAWD models caused by the accretion and Roche-lobe mass loss. 
    } 
    \label{fig:fig8}
\end{figure}

The new computations of the RAWD multicycle evolution confirm the
conclusion about their low mass retention efficiency (the last column
of Table~\ref{tab:models} and Figure~\ref{fig:fig8}) made in Paper~I,
at least for [Fe/H]\,$\ga -2$.  A new result is that $\eta$
increases when [Fe/H] decreases below $-2$ to
fractions of $\approx 20 $ to $30\%$ in models E, F and
G. The $0.75 M_\odot$ RAWD model
with [Fe/H]\,$=-2.6$ (model G) has the highest value of $\eta = 29\%$ and a
lower He peak luminosity of $\log_{10}L_\mathrm{He}^\mathrm{max} =
8.5$ (the last row in Table~\ref{tab:models}) compared to models E and
F. According to these models the SD channel of SNIa progenitors may still
work at very low metallicities.

The revealed trend of $\eta$ with [Fe/H] is probably caused by the
lower opacity of the metal-poor accreted matter. It allows the
accumulated He layer to cool down, and therefore be compressed by the
gravity more efficiently. As a result, the He flash starts at a lower
mass of the He shell and achieves a lower peak luminosity, which is
reflected in the different amplitudes of the RAWD mass changes in the
models with [Fe/H]\,$<-1.55$ (E and F) compared to the models with
[Fe/H]\,$\geq -1.55$ (B, C and D) in Figure~\ref{fig:fig8}, and in
their different values of
$\log_{10}L_\mathrm{He}^\mathrm{max}/L_\odot$
(Table~\ref{tab:models}). Besides, during the expansion of the RAWD
following the He-shell flash its envelope cools down faster, because
of the lower opacity, in the low-metallicity models, and the RAWD
returns to the accretion phase after having lost a smaller fraction of
the accreted matter.

As for the RAWD models with [Fe/H]\,$\geq -1.55$, they stubbornly want
to expand to red-giant dimensions and it is only because of our
implementation of the Roche-lobe mass loss that they can expand only
up to $R\approx R_\mathrm{RL,WD}$ (Figure~\ref{fig:fig2}a) and
retain this radius untill a significant fraction of the accreted
matter is gone with the wind.  To test if this behaviour is affected
by our choice of the \mesa\ mass-loss algorithm, we have switched to
the super-Eddington wind prescription in model C, enforcing it to work
only when $R\approx R_\mathrm{RL,WD}$.  With this modification, we
have reproduced the results obtained for model C with the Roche-lobe
mass-loss prescription.

\begin{figure}
  \includegraphics[width=\columnwidth]{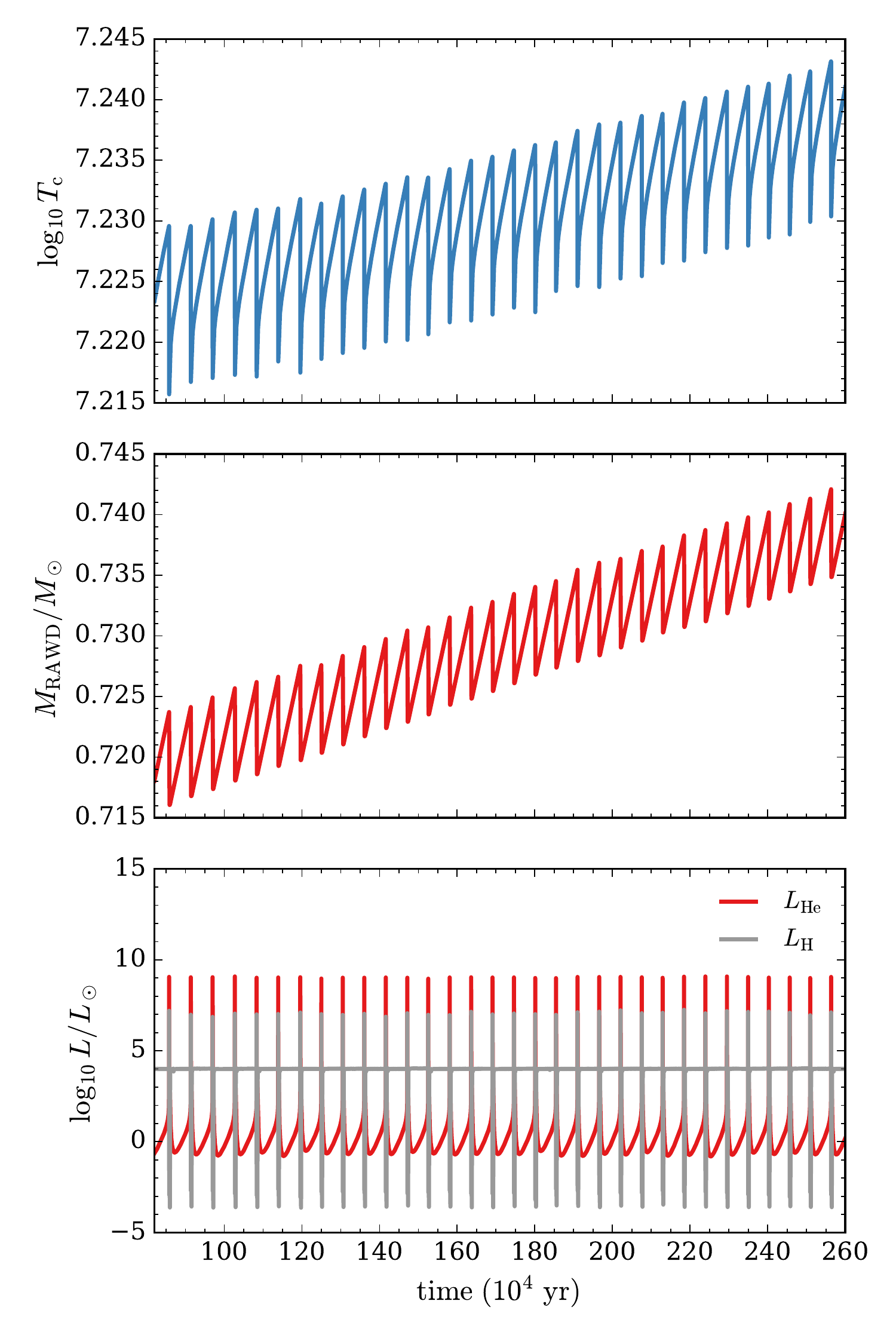}
  \caption{Central temperature, total RAWD mass and H- and
      He-burning luminosity as a function of time for the long-term
    multicycle evolution of model C.}
    \label{fig:fig9}
\end{figure}

We have also addressed the frequently raised concern that a much
larger number of He-shell flashes than we have simulated in our RAWD
models could lead to significant changes in the RAWD multicycle
evolution. To test this, we have extended the computed number of the
He-shell flashes in model C up to 42 from the initial 10. The results
of this long-run simulations are plotted in
Figure~\ref{fig:fig9}. Its top panel demonstrates that the WD
central temperature has increased only by $6\%$.  Our analysis of the
RAWD mass variations presented in the middle panel shows that the mass
retention efficiency has varied between $5.8\%$ and $8.8\%$ in this
run, still remaining below 10\%, as it was in the initial 10 cycles. 
Finally, the bottom panel
reveals that the He peak luminosity has not changed at all.

\subsection{The RAWD i-process nucleosynthesis yields}
\label{sec:yields}

Before presenting the yields, we will summarize how we choose the input
data for the RAWD i-process nucleosynthesis computations. For this
purpose, we use model C (Figure~\ref{fig:fig2}), as an example.  We
begin with selecting a representative RAWD
He-shell flash that does not stands out, therefore this cannot be the
first He-shell flash that usually is stronger than the others. The
only exception to this rule in the present paper is the
solar-metallicity model A. We find it to be much more difficult to
compute multiple cycles with stable H burning interrupted by strong
He-shell flashes at a high metallicity, probably because of the high
opacity of the envelope matter. Because the multicycle evolution of
the solar-metallicity RAWD models has already been discussed in
Paper~I, and the i-process yields calculated for the first He-shell
flash in the solar-metallicity RAWD model in this work are similar to
those presented in Paper~I, we have followed only one He-shell flash
in the new model A. As for model C, we have chosen the 6th flash
(panels a, b and c in Figure~\ref{fig:fig2}). We have used its
corresponding H-burning luminosity (panel d) to estimate the
H-ingestion rate with Equation~\ref{eq:dMHdt} (the purple curve in
panel e).  The dashed line in panel e is our estimate of an average
$\dot{M}_\mathrm{H}$ value for this model that was used to calculate
the parameter $\dot{M}_\mathrm{ing} =
(\dot{M}_\mathrm{H}/X_\mathrm{surf})$ listed in
Table~\ref{tab:models}. The vertical dotted lines
in panel e constrain $t_\mathrm{ing}$. The left line is chosen
close to the beginning of H ingestion. The position of the right line
is less certain. It marks the beginning of the fast decline of
$\dot{M}_\mathrm{H}$.
The H-ingestion time in the RAWD models is usually less than one month
($t_\mathrm{ing}\la 0.08\ \mathrm{yr}$), except the first phase of
H-ingestion in model A. Like in the solar-metallicity model A from
Paper~I, the updated model A has two phases of H ingestion, the
longer-lasting slow-ingestion phase with $\dot{M}_\mathrm{ing} =
2.2\times 10^{-12}\ M_\odot\ \mathrm{s}^{-1}$ and $t_\mathrm{ing} =
0.17$ yr followed by the shorter fast-ingestion phase with
$\dot{M}_\mathrm{ing} = 3.5\times 10^{-11}\ M_\odot\ \mathrm{s}^{-1}$
and $t_\mathrm{ing} = 0.024$ yr, both included in our nuclesosynthesis
computations.

It usually takes about hundred time steps for the \mesa\ code to
evolve a RAWD model through the entire H-ingestion phase, meaning that
the time interval between two consecutive models on this evolutionary
phase is hundreds of minutes, corresponding to tens of convective
turn-over times of the He-shell flash convection.
This is too long for the i-process
nucleosynthesis simulations that usually require a time step of the
order of minutes \citep{Herwig2011}.
We cannot reduce the \mesa\ time steps, because the mixing-length theory (MLT)
adopted in \mesa\ to describe convection is formulated in terms of
time and spatial averages. Time steps smaller than about ten times
the convective turnover time would violate this assumption.
Therefore, we simply take the temperature, density, radius and MLT
convective diffusion coefficient profiles from a \mesa\ model in the
middle of the H-ingestion phase, when the accreted envelope is already
expanding (e.g., the dashed $T$ profile in panel f), and use them to
set up our post-processing nucleosynthesis simulations.
Note, that \cite{Herwig2011} were able to reproduce
the surface abundances of heavy elements measured in Sakurai's object
by \cite{Asplund1999} using a 1D model similar to this one.

\begin{figure}
  \includegraphics[width=\columnwidth]{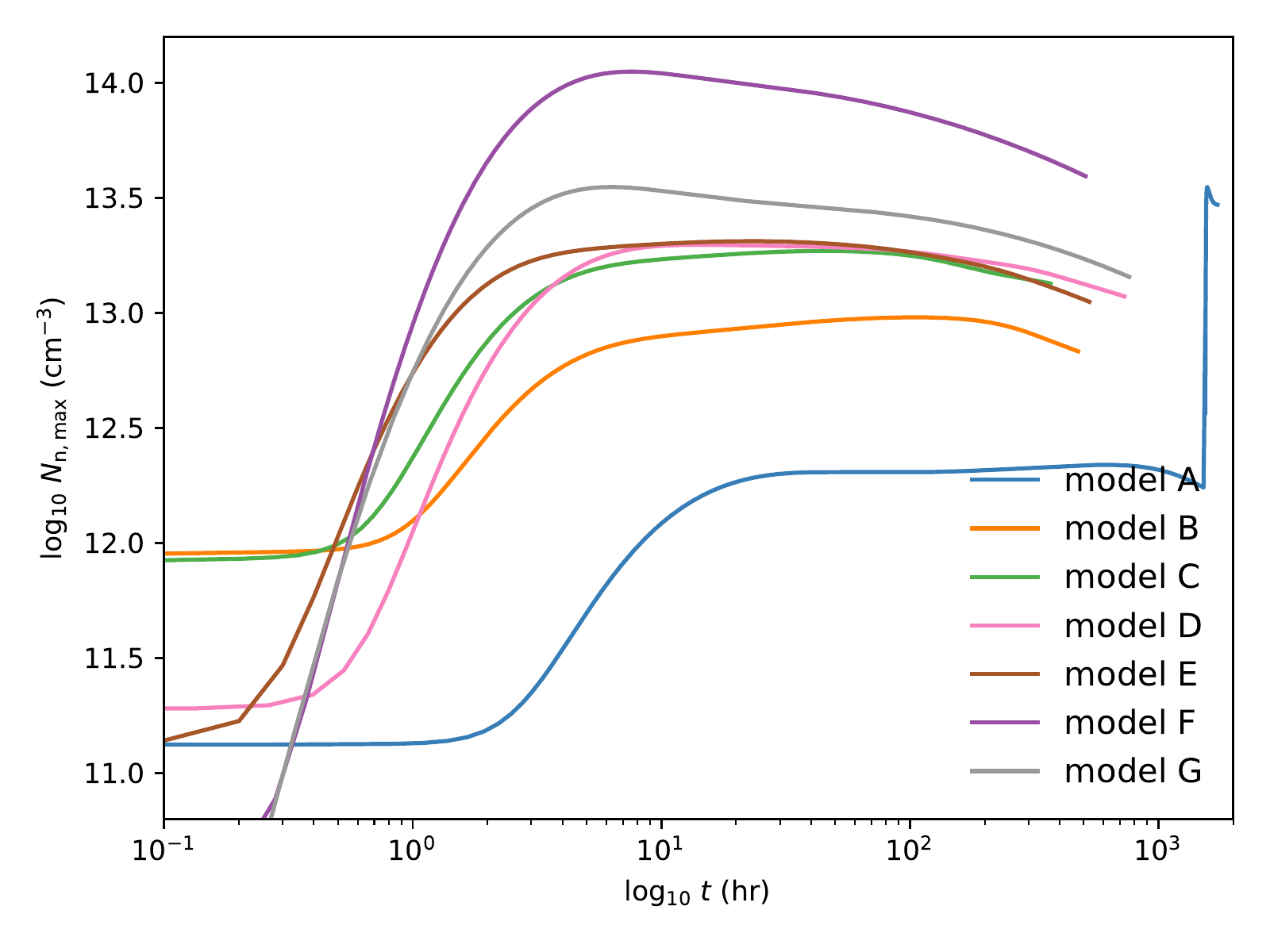}
  \caption{The evolution of the maximum neutron number density in the He convective zones of
           our RAWD models. In model A, the jump in the evolution of $N_\mathrm{n,max}$ at
           the end is caused by the switching to the second phase of H ingestion that is shorter but faster than
           the previous phase (see text). 
    } 
    \label{fig:fig10}
\end{figure}

\begin{figure}
  \includegraphics[width=\columnwidth]{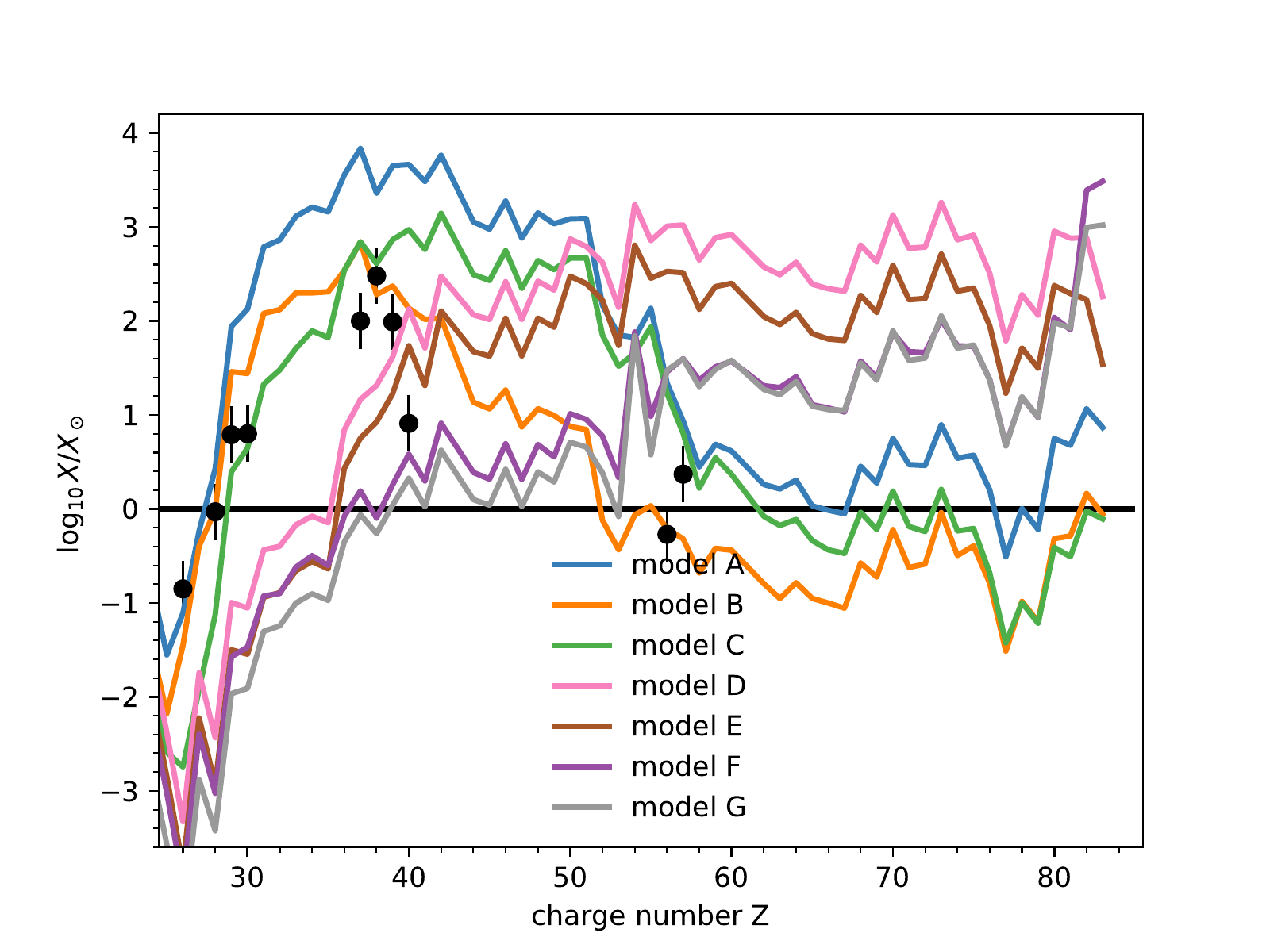}
  \caption{The elemental i-process yields (solar-scaled mass fractions) from our RAWD models. For comparison,
           the abundances of the first peak n-capture elements measured in Sakurai's object by
           \protect\cite{Asplund1999} are shown as filled black circles with errorbars. They
           were interpreted by \protect\cite{Herwig2011} as results of the i-process nucleosynthesis in the convective
           He shell during its very last thermal pulse in a model of Sakurai's object with a half-solar metallicity.
    } 
    \label{fig:fig11}
\end{figure}

The post-processing simulations of the i-process nucleosynthesis were
carried out using the methods described in
\Sect{ipr_meth}.
The nucleosynthesis simulations provide the abundance
distributions in the He convective zone and the surface
abundances at the top
of the He shell for the selected ($\sim 1000$) isotopes for $0\leq
t\leq t_\mathrm{ing}$. At the end, we allow the surface abundances to
decay for 1 Gyr. Figure~\ref{fig:fig10} shows the evolution of the
maximum neutron number density in the He shell for our RAWD
models. The peak value of the $N_\mathrm{n,max}(t)$ curves increases
with a decrease of [Fe/H] because the total mass fraction of the
isotopes that capture neutrons decreases with the metallicity, while
the production of neutrons, that is controlled by the $^{12}$C
abundance in the He shell and the total amount of ingested H, remains
approximately the same. In other words, the neutron source is primary, 
while the i-process seeds are secondary. The steep increase of $N_\mathrm{n,max}$ at
the end of its evolution in model A marks the beginning of the short
fast-ingestion phase that appears to be common for solar-metallicity
RAWDs (Paper~I).  It demonstrates that $N_\mathrm{n,max}$ increases
with $\dot{M}_\mathrm{ing}$ for models with the same metallicity.

Figure~\ref{fig:fig11} shows the final surface elemental abundances
divided by the solar abundances from \cite{Asplund2009} of the RAWD
models\footnote{The files with the elemental and isotopic abundance
yields from our RAWD models can be found at
\url{http://apps.canfar.net/storage/list/nugrid/data/projects/RAWD/iRAWDyields}}.
As in the \sprn\ the global heavy-element distribution shifts to
higher mass elements at lower metallicity
\citep{clayton:68,busso:01a}. 
We therefore, generally speaking, expect to see
local elemental i-process signatures in higher-mass second- and
third-peak species at lower metallicity, whereas the i-process signature
may be most prominently detected in lower-mass, first-peak elements at
higher, solar-like metallicities. Accordingly it had been proposed in
Paper I that the solar-metallicity RAWDs could be contributors of
first-peak elements to the solar system abundance
distribution. \cite{Cote2018} have used our RAWD i-process yields in a
framework that included Galactic chemical evolution and binary-star
population synthesis models to confirm this hypothesis.

An example for solar-like metallicity i-process abundances are those of
Sakurai's object which indeed shows large enhancements in the first
peak n-capture elements (around $Z=40$) as shown in Figure~\ref{fig:fig11}. The
details of the abundance patterns of Sakurai's object are however better
reproduced by the very-late thermal pulse post-AGB star models of
\citet{Herwig2011} which feature peak neutron densities and H
ingestion rates that are both more than two orders of magnitude
higher compared to the RAWD models.

\subsection{A possible relation of the metal-poor RAWDs to the CEMP-r/s stars}

An example for i-process abundance patterns in the second-peak
elements expected at lower metallicity may be found in the
subclass of carbon-enhanced metal-poor (CEMP) stars
with abundance patterns that appear to be enhanced with both
$r$- and \spr\ elements
\citep[e.g.][]{Beers2005,Masseron2010,Lugaro2012,Bisterzo2012}.
\cite{Dardelet2015} and \cite{Hampel2016} have done one-zone
nucleosynthesis simulations of i-process conditions to
demonstrate that the abundances of heavy elements observed in these
CEMP-r/s stars can be reproduced by an n-capture process with neutron
densities $N_\mathrm{n} = 10^{12}$\,--\,$10^{15}\ \mathrm{cm}^{-3}$.

\begin{figure}
  \includegraphics[width=\columnwidth]{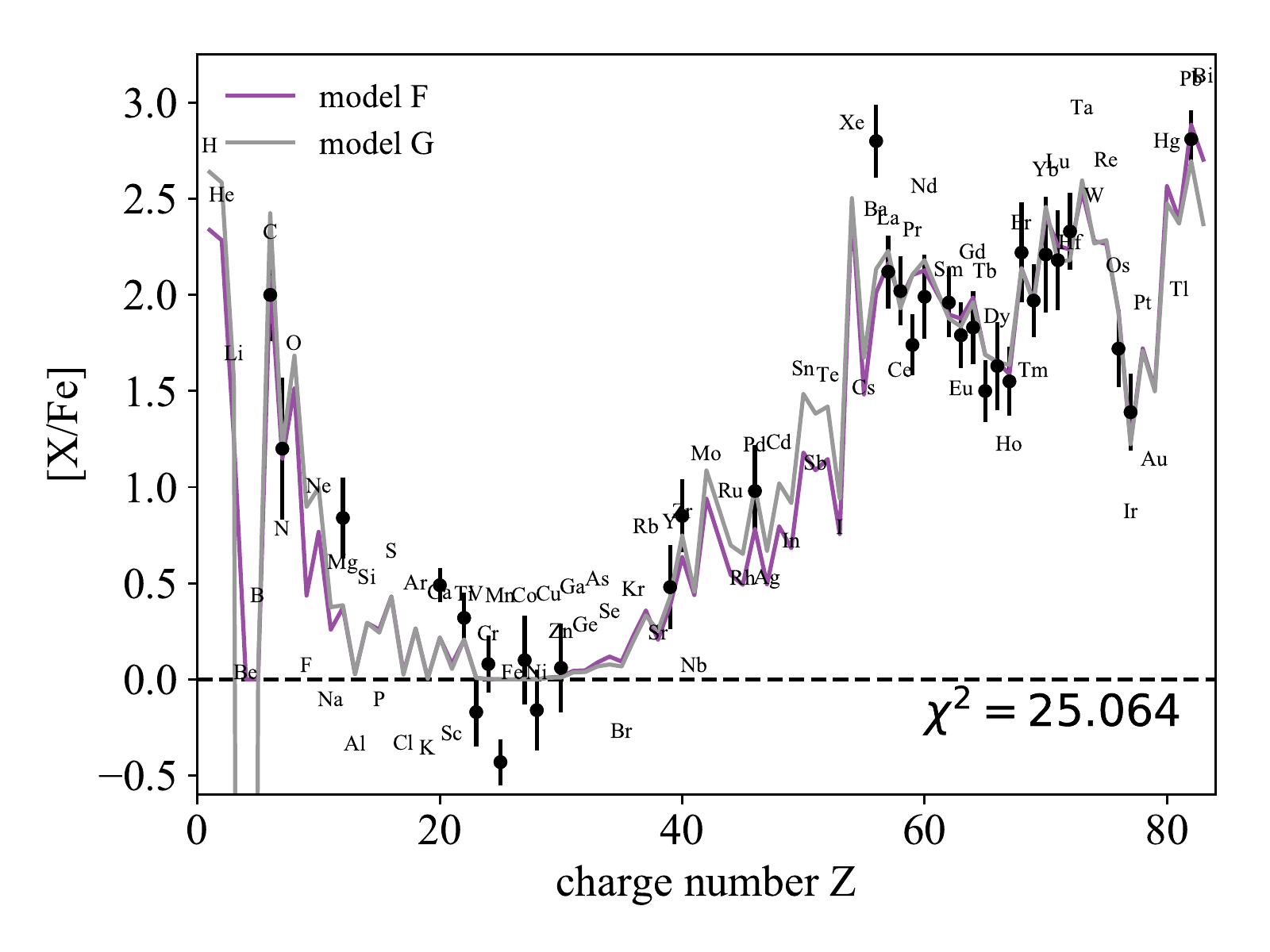}
  \caption{Abundances of heavy elements observed in the CEMP-r/s star
    CS31062-050 \citep{Aoki2002,Johnson2004} and the best-fit
      abundance distribution from the time evolution of the
      RAWD model G, that has [Fe/H]\,$=-2.6$, diluted with 99.58\% of the initial abundances.
      Most CEMP-r/s stars have metallicites in the range of $-3\leq\mathrm{[Fe/H]}\leq -2$
       \citep[e.g.][]{Abate2016}.
       A similarly good fit for the star CS31062-050 is obtained with the RAWD model F.
      The element names are placed either above or below (so that they do not overlap)
      the predicted abundances (curves). 
      }
    \label{fig:fig12}
\end{figure}

As an example, Figure~\ref{fig:fig12} shows the best $\chi^2$ fit of
the i-process elemental yields from our RAWD models G and F to the surface
chemical composition of the CEMP-r/s star CS31062-050 that has the metallicity
[Fe/H]\,$=-2.42$ \citep{Johnson2004} intermediate between those assumed for
our models F and G.
The only exception to the otherwise excellent agreement is the
discrepant Ba abundance that requires a further
investigation. For other
CEMP-r/s stars a good agreement with the RAWD i-process yields can be found as
well, including their lower Ba abundances. The low-metallicity RAWD models are at this
point the first and only models in which nucleosynthesis calculation
directly post-processing complete stellar evolution models can
reproduce the complete abundance patterns observed in CEMP-r/s
stars. We therefore propose that CEMP-r/s stars that have been well
reproduced with i-process models should be refered to as CEMP-i stars.

Our findings suggest a new scenario for the formation of CEMP-r/s, or
in this case CEMP-i stars, that takes into account our finding that
the RAWDs with [Fe/H]\,$\la -2$ may reach the Chandrasekhar mass and
explode as SNe Ia. It is based on the fact that the mass retention
efficiency of our RAWD models significantly increases when [Fe/H]
decreases below $-2$ (the last column of Table~\ref{tab:models}), and
it is supported by our calculations of the evolution of binary-star
parameters. 
%results of which are presented in
%Figures~\ref{fig:binRAWD} and \ref{fig:binSNIa}. 
For these
calculations, we have used the isotropic re-emission model of mass
transfer in which a fraction $\beta$ of matter accreted by the primary
star (RAWD) is lost from the binary system. In our case, $\beta =
1-\eta$, where $\eta$ is the mass retention efficiency.  This model
and its equations are described by \cite{Postnov2014} in their Section
3.3.3.  
%Our Figures~\ref{fig:binRAWD} and \ref{fig:binSNIa} show the
We have solved these equations to model the
evolution of the semi-major axis $a$, the mass ratio $q = M_1/M_2$,
and the RAWD mass $M_1$ for two sets of the binary initial parameters.
%Figure~\ref{fig:binRAWD} corresponds to the 
In both cases we start with $M_2 = 2.5\,M_\odot$ and use $\dot{M}_2 = -1.5\times 10^{-7} M_\odot\,\mathrm{yr}^{-1}$.
The first case assumes that $\eta = 10\%$ which
includes the RAWD models with [Fe/H]\,$\ga -2$, while
%Figure~\ref{fig:binSNIa} 
the second case has $\eta = 30\%$ and demonstrates a possible evolution of the
binary-system parameters for RAWDs with [Fe/H]\,$\la -2$. In both
cases, we have stopped the calculations at $q=2$.

In the first case, $M_1$ grows from $0.75\,M_\odot$ to only $\sim 0.9\,M_\odot$, while
in the second case $M_1$ starts growing from $0.85\,M_\odot$ and it does reach the Chandrasekhar limit, and the RAWD
ends its life as a SNIa. If this is true, then some of the
present-day CEMP-i stars could be former tertiary members of hierarchical triple
systems in which they had been orbiting a close binary system with a
RAWD. A series of dozens He-shell flashes on the RAWD, each being
followed by the RAWD expansion and mass loss, could enrich the
tertiary star with the products of i-process nucleosynthesis.
This enrichment scenario is similar to the one proposed to explain
abundace anomalies in CEMP-s stars by accretion of
material lost by their AGB star binary companions \citep[e.g.,][and references therein]{Abate2018}.
A difference is that tertiary stars in hierarchical triple systems are
farther away from their polluting primary components. This should result in
a stronger dilution of the accreted material. But, on the other hand,
heavy elements produced in the s process in AGB stars are already diluted
in their relatively massive envelopes before they are ejected, while
the dilution in a thin envelope of an RAWD is negligible. If thermohaline mixing
in envelopes of accreting stars leads to an even stronger dilution
depends on the efficiency (timescale) of this mixing which still remains uncertain \citep{Denissenkov2008,Stancliffe2007}.
For example, thermohaline mixing can be suppressed by strong horizontal turbulent diffusion \citep{Denissenkov2010}.
Finally, when the RAWD exploded as SNIa, the tertiary star, that
became a CEMP-i star by that moment, would leave the triple
system because of a decreased gravitational pull to the center of mass 
that has suddenly lost $\sim 1.4\,M_\odot$. Those CEMP-i stars would not be binaries anymore. If the RAWD
does not explode as a SN Ia the CEMP-i star would be possibly in a
wider orbit around compact binary and would show a long binary period
super imposed with the very short period of the compact RAWD
binary. Thus, CEMP-i stars can be both single stars and binaries in
this scenario. Our scenario is supported by the recent finding that
the fractions of binary and triple star systems in stellar populations significantly increase
with a decreasing metallicity \citep{Fuhrmann2017,Badenes2018,Moe2019}.
 
This scenario can potentially explain the observational bias against
finding CEMP-r/s stars in globular clusters that may be caused by the
destruction of wide triple systems through close star encounters in
dense cores of globular clusters. Our scenario differs from the
triple-system scenario for the formation of CEMP-r/s stars discussed
by \cite{Abate2016}, in which a primary massive star was assumed to
produce the \rpr\ elemental abundances during its SN explosion and
the presence of a secondary AGB star was required to make \spr\
elements. The plausibilty of our scenario could be checked with
triple-star population synthesis simulations, but this is out of the scope of the present paper. 

%\begin{figure}
%  \includegraphics[width=1.05\columnwidth]{figs/binary_evol_RAWD.pdf}
%  \caption{The evolution of the binary-system parameters, the semi-major axis $a$, the mass ratio $q = M_1/M_2$, 
%           and the RAWD mass $M_1$, for the isotropic re-emission model of mass transfer in which
%           $\beta = 1-\eta$, where $\eta$ is the RAWD mass retention efficiency. This set of the initial
%           binary parameters corresponds to the RAWD models with [Fe/H]\,$\ga -2$ that have $\eta\la 10\%$.
%           The calculations stop when $q=2$.
%    } 
%    \label{fig:binRAWD}
%\end{figure}

%\begin{figure}
%  \includegraphics[width=1.05\columnwidth]{figs/binary_evol_SNIa.pdf}
%  \caption{Same as in Figure~\ref{fig:binRAWD}, but for the RAWD models with [Fe/H]\,$\la -2$ that have
%           $\eta\ga 20\%$. In this case that assumes $\eta = 30\%$, the RAWD mass may reach the Chandrasekhar limit.
%    } 
%    \label{fig:binSNIa}
%\end{figure}

\section{Summary}
\label{sec:concl}

We have used the \mesa\ stellar evolution code \citep{Paxton2011,Paxton2013} to compute the models of CO white dwarfs (WDs)
rapidly accreting H-rich matter, assuming this matter is donated by normal (main-sequence, sub-giant or red-giant-branch)
components of the WDs in close binary systems. Such stellar configurations can result from
common envelope events, after the primary AGB star components fill their Roche lobes, loose almost entire
envelopes above the WD cores in unstable mass transfers, and the binary systems become tighter
via the transformation of their orbital energies into the kinetic energy  of the ejecta. When the secondary
components of these post-common-envelope systems expand and fill their own Roche lobes by or after the end of
the main-sequence evolution, they begin to donate H-rich matter to their, by this time cooled-down, WD          
primary components. We assume that the mass accretion rate is rapid enough, $M_\mathrm{acc}\sim 10^{-7}\ M_\odot\ \mathrm{yr}^{-1}$,
for the accreted H to be stably burning on the WD surface, resulting in the accumulation of a He shell.
When its mass reaches a critical value, the He shell will experience a thermal flash in which some fraction of
He will be transformed into C. If the sequence of the stable H burning intermittent with the He-shell flashes
is not accompanied by a significant mass loss by the WD, then its mass may eventually reach the Chandrasekhar limit,
$M_\mathrm{Ch}\approx 1.38 M_\odot$, and the CO WD will explode as a SNIa. This is the classical single degenerate (SD)
channel of SNIa progenitors \citep{Schatzman1963,Whelan1973}.

In this work, the evolution of such rapidly-accreting WDs (RAWDs) has for the first time been followed through
multiple strong He-shell flashes. Each He-shell flash leads to the expansion of the WD envelope that overflows
the WD Roche lobe. We assume that the matter leaving the WD Roche lobe is lost from the binary system because of
its interaction with the secondary component, like in the common-envelope event. The important new result found
in our simulations is that the WD envelope remains inflated at the WD Roche-lobe radius until its significant
fraction is lost, after which it shrinks, and the mass accretion onto the WD resumes. As a result, we have
obtained relatively low mass retention efficiencies, $\eta \la 10\%$, for our RAWD models, at least for
the metallicities [Fe/H]\,$\ga -2$ (Table~\ref{tab:models}). It is unlikely that the masses of such RAWDs will ever approach the Chandrasekhar
limit. Therefore, the SD channel should not work in these cases. As for the RAWD models with [Fe/H]\,$\la -2$,
they are found to have $\eta \ga 20\%$ and, therefore, we cannot exclude that their masses will ultimately reach
the Chandrasekhar limit.

The ignition of He at the bottom of the He shell triggers convection. When the top of the He-flash convective zone
reaches the bottom of the H-rich envelope, convective boundary mixing (convective shear mixing and convective overshooting)
starts to ingest protons into the He shell. There, they are quickly captured by the abundant $^{12}$C nuclei producing
the unstable $^{13}$N that has the lifetime of $\sim 10$ minutes and decays into $^{13}$C while being transported
by convection downwards. When $^{13}$C nuclei arrive at the bottom of the He shell they capture $\alpha$ particles releasing
neutrons. This convective-reactive process transforms almost every proton ingested at the top of the He shell
into a neutron at its bottom, therefore at sufficiently high mass ingestion rates, 
$M_\mathrm{ing}\sim 10^{-12}$--$10^{-11}\ M_\odot\ \mathrm{s}^{-1}$, the neutron density can be as high as
$N_\mathrm{n,max}\sim 10^{12}$\,--\,$10^{15}\ \mathrm{cm}^{-3}$. These values are intermediate between those
characteristic of the $s$ and $r$ processes, therefore the ensuing n-capture process is called the i-process \citep{Cowan1977}.

The important input parameters ---
the ingestion rate and convective boundary mixing efficiency --- have been
determined and constrained through a series of three-dimensional
hydrodynamic simulations of the RAWD He-shell flash convection. The
estimates of the mass ingestion rate obtained from
1D RAWD models are consistent with the 3D hydrodynamic simulations. The convective boundary mixing efficiency parameter adopted in our stellar evolution simulations at the top of the He-shell flash convection boundary are in agreement with the luminosity scaling law for the CBM  $f_\mathrm{top}$ parameter presented here.

It is interesting that the i-process nucleosynthesis yields predicted
by our metal-poor ([Fe/H]\,$\la -2$) RAWD models almost perfectly fit
the abundances of heavy elements measured in some CEMP-r/s stars (Figure~\ref{fig:fig12}). 
These are the CEMP-i stars. Given that the same RAWD models have the
higher mass retention efficiencies and can potentially become SNeIa,
we propose that CEMP-i stars used to be distant members
of triple star systems orbiting close binary systems with
RAWDs. When the RAWD exploded as a SNIa, the tertiary star, polluted by
the products of i-process nucleosynthesis that had taken place on the
RAWD, got loose from the system and is now seen as a single star. If
the RAWD has not exploded yet, the CEMP-i star can still be a member
of a triple system, and would then show signs of binarity.

\section*{Acknowledgements}

FH acknowledges funding from NSERC through a Discovery Grant. This
research is supported by the National Science Foundation (USA) under
Grant No. PHY-1430152 (JINA Center for the Evolution of the Elements).

%%%%%%%%%%%%%%%%%%%%%%%%%%%%%%%%%%%%%%%%%%%%%%%%%%

%%%%%%%%%%%%%%%%%%%% REFERENCES %%%%%%%%%%%%%%%%%%

% The best way to enter references is to use BibTeX:

\bibliographystyle{mnras}
\bibliography{paper.bib} % if your bibtex file is
                                             % called example.bib

%%%%%%%%%%%%%%%%%%%%%%%%%%%%%%%%%%%%%%%%%%%%%%%%%%

%%%%%%%%%%%%%%%%% APPENDICES %%%%%%%%%%%%%%%%%%%%%

\appendix

\section{The implementation of mass ingestion in RAWD models}
\label{app:mdot}

Let us denote $X_\mathrm{i}^k$ and $X_\mathrm{o}^k$
the mass fractions of the $k$th isotope, respectively, inside and
outside the convective He shell, in the vicinity of its outer
boundary.  We assume that the envelope matter ingested into the He
shell gets immediately distributed within an ingestion zone with the
mass $\Delta M\ll M_\mathrm{top}-M_\mathrm{bot}$, where
$M_\mathrm{bot}$ and $M_\mathrm{top}$ are the mass coordinates of the
bottom and the top of the He shell. The input parameters here are $X_\mathrm{o}^k$,
$\Delta M$, and the mass ingestion rate $\dot{M}_\mathrm{ing}$.
Because of a small size of the
ingestion zone, we assume that all the isotopes are linearly
distributed in it, i.e.
$$
X_\mathrm{i}^k(M) = X_\mathrm{i}^k(M_\mathrm{max}) - \frac{(M_\mathrm{max}-M)}{\Delta M}
[X_\mathrm{i}^k(M_\mathrm{max})-X_\mathrm{i}^k(M_\mathrm{min})],
$$
where $M_\mathrm{min}\leq M\leq M_\mathrm{max}$, $M_\mathrm{max} = M_\mathrm{top}$, and $M_\mathrm{min} = M_\mathrm{max}-\Delta M$.
The total mass of the $k$th isotope in this distribution is
$$
\Delta M_\mathrm{i}^k = \int_{M_\mathrm{min}}^{M_\mathrm{max}} X_\mathrm{i}^k(M)dM =
\frac{1}{2}[X_\mathrm{i}^k(M_\mathrm{min}) + X_\mathrm{i}^k(M_\mathrm{max})]\Delta M.
$$
After $\Delta M_\mathrm{ing}=\dot{M}_\mathrm{ing}\Delta t$ of the envelope matter is ingested, the mass of the $k$th isotope in the ingestion zone becomes
$$
\widetilde{\Delta M_\mathrm{i}^k} = \Delta M_\mathrm{i}^k + X_\mathrm{o}^k\Delta M_\mathrm{ing} - \delta M_\mathrm{i}^k,
$$
where
\begin{eqnarray}
\nonumber
\delta M_\mathrm{i}^k = \int_{M_\mathrm{max}-\Delta M_\mathrm{ing}}^{M_\mathrm{max}} X_\mathrm{i}^k(M)dM & = &
X_\mathrm{i}^k(M_\mathrm{max})\Delta M_\mathrm{ing} \\ \nonumber
+ \frac{1}{2}\frac{\Delta M_\mathrm{ing}^2}{\Delta M}[X_\mathrm{i}^k(M_\mathrm{min}) & - &
X_\mathrm{i}^k(M_\mathrm{max})]
\end{eqnarray}
is the mass of the $k$th isotope in the mass $\Delta M_\mathrm{ing}$ that replaces the ingested mass outside the He shell
(we assume that the mixing between the He shell and the envelope does not change the chemical composition, $X_\mathrm{o}^k$, of the latter.)

The change of the average mass fraction of the $k$th isotope in the ingestion zone is therefore
\begin{eqnarray}
\nonumber
\Delta \langle X_\mathrm{i}^k\rangle & = & \frac{\widetilde{\Delta M_\mathrm{i}^k}}{\Delta M} -
\frac{\Delta M_\mathrm{i}^k}{\Delta M} = [X_\mathrm{o}^k - X_\mathrm{i}^k(M_\mathrm{max})]\frac{\Delta M_\mathrm{ing}}{\Delta M} \\ \nonumber
& + & \frac{1}{2}[X_\mathrm{i}^k(M_\mathrm{max}) - X_\mathrm{i}^k(M_\mathrm{min})]\left(\frac{\Delta M_\mathrm{ing}}{\Delta M}\right)^2.
\end{eqnarray}
This change can be applied either as a step increase of the previous
linear distribution
$$
\widetilde{X_\mathrm{i}^k(M)} = X_\mathrm{i}^k(M) + \Delta \langle X_\mathrm{i}^k\rangle,
$$
or as a ramp increase
$$
\widetilde{X_\mathrm{i}^k(M)} = X_\mathrm{i}^k(M) + 2\frac{(M-M_\mathrm{min})}{\Delta M}\Delta \langle X_\mathrm{i}^k\rangle
$$
for $M_\mathrm{min}\leq M\leq M_\mathrm{max}$.
In the present work, we have chosen the latter option.

%%%%%%%%%%%%%%%%%%%%%%%%%%%%%%%%%%%%%%%%%%%%%%%%%%

% Don't change these lines
\bsp	% typesetting comment
\label{lastpage}
\end{document}